\documentclass[jimaging,article,accept,pdftex,moreauthors]{Definitions/mdpi} 
\firstpage{1} 
\pubvolume{8}
\issuenum{10}
\articlenumber{259}
\pubyear{2022}
\copyrightyear{2022}
\externaleditor{Academic Editor: Silvia Liberata Ullo 
}
\datereceived{10 August 2022} 
\dateaccepted{16 September 2022} 
\datepublished{22 September 2022} 
\hreflink{https://doi.org/10.3390/jimaging8100259} 

\usepackage[labelformat=simple]{subcaption}

\DeclareCaptionLabelFormat{subcaptionlabel}{\normalfont(\textbf{#2}\normalfont)}
\usepackage{pgfplotstable}
\usepackage{pgfplots}
\usepackage{textcomp}
\usepackage{amsfonts,amssymb,bbm}	
\usepackage{multicol}
\usepackage{tabularx} 

\usepackage[section]{placeins}

\Title{DS6, Deformation-Aware Semi-Supervised Learning: Application to Small Vessel Segmentation with Noisy \linebreak Training Data}

\TitleCitation{DS6---Deformation-Aware Semi-Supervised Learning: Application to Small Vessel Segmentation with Noisy Training Data}

\Author{Soumick Chatterjee 
 $^{1,2,3,}$*\orcidA{}, Kartik Prabhu $^{1,\dagger}$, Mahantesh Pattadkal $^{1,\dagger}$, Gerda Bortsova $^{4}$, \linebreak Chompunuch Sarasaen $^{3,5}$, Florian Dubost $^{4}$, Hendrik Mattern $^{3}$, Marleen de Bruijne 
 $^{4,6}$, Oliver Speck $^{3,7,8}$ \linebreak and Andreas N{\"u}rnberger $^{1,2,8}$}

\AuthorNames{Soumick Chatterjee, Kartik Prabhu, Mahantesh Pattadkal, Gerda Bortsova, Chompunuch Sarasaen, Florian Dubost, Hendrik Mattern, Marleen de Bruijne, Oliver Speck and Andreas N{\"u}rnberger }

\AuthorCitation{Chatterjee, S.; 
 Prabhu, K.; Pattadkal, M.; Bortsova, G.; Sarasaen, C.; Dubost, F.; Mattern, H.; de Bruijne, M.; Speck, O.; N{\"u}rnberger, A.}

\address{%
$^{1}$ \quad Faculty of Computer Science, Otto von Guericke University Magdeburg, 39106 Magdeburg, Germany; 
 \linebreak kartik.prabhu@ovgu.de (K.P.); mahantesh.pattadkal@ovgu.de (M.P.); andreas.nuernberger@ovgu.de (A.N.)\\
$^{2}$ \quad Data and Knowledge Engineering Group, Otto von Guericke University, 39106 Magdeburg, 
 Germany\\
$^{3}$ \quad Biomedical Magnetic Resonance, Otto von Guericke University Magdeburg, 39106 Magdeburg, Germany; chompunuch.sarasaen@ovgu.de (C.S.); hendrik.mattern@ovgu.de (H.M.); oliver.speck@ovgu.de (O.S.)\\
$^{4}$ \quad Biomedical Imaging Group Rotterdam, Department of Radiology \& Nuclear Medicine, Erasmus MC, \linebreak 3015 GD Rotterdam, The Netherlands; gerdabortsova@gmail.com (G.B.); floriandubost1@gmail.com (F.D.); marleen.debruijne@erasmusmc.nl (M.d.B.)\\
$^{5}$ \quad Institute for Medical Engineering, Otto von Guericke University Magdeburg, 39106 Magdeburg, Germany\\
$^{6}$ \quad Department of Computer Science, University of Copenhagen, DK-2100 Copenhagen, Denmark\\ 
$^{7}$ \quad German Center for Neurodegenerative Disease, 39120 Magdeburg, Germany\\
$^{8}$ \quad Center for Behavioral Brain Sciences, 39106 Magdeburg, Germany}

\corres{Correspondence: soumick.chatterjee@ovgu.de}

\firstnote{These authors contributed equally to this work.} 

\abstract{Blood vessels of the brain provide the human brain with the required nutrients and oxygen. As a vulnerable part of the cerebral blood supply, pathology of small vessels can cause serious problems such as Cerebral Small Vessel Diseases (CSVD). It has also been shown that CSVD is related to neurodegeneration, such as Alzheimer's disease. With the advancement of 7 Tesla MRI systems, higher spatial image resolution can be achieved, enabling the depiction of very small vessels in the brain. Non-Deep Learning-based
 approaches for vessel segmentation, e.g., Frangi's vessel enhancement with subsequent thresholding, are capable of segmenting medium to large vessels but often fail to segment small vessels. The sensitivity of these methods to small vessels can be increased by extensive parameter tuning or by manual corrections, albeit making them time-consuming, laborious, and not feasible for larger datasets. This paper proposes a deep learning architecture to automatically segment small vessels in 7 Tesla 3D Time-of-Flight (ToF) Magnetic Resonance Angiography (MRA) data. The algorithm was trained and evaluated on a small imperfect semi-automatically segmented dataset of only 11 subjects; using six for training, two for validation, and three for testing. 
 The deep learning model based on U-Net Multi-Scale Supervision was trained using the training subset and was
  made equivariant to elastic deformations in a self-supervised manner using deformation-aware learning to improve the generalisation performance. The proposed technique was evaluated quantitatively and qualitatively against the test set and achieved a Dice score of 80.44 $\pm$ 0.83. Furthermore, the result of the proposed method was compared against a selected manually segmented region (62.07 resultant Dice) and has shown a considerable improvement (18.98\%) with deformation-aware learning. }


\begin{document}

\section{Introduction}
Small vessels in the brain, such as the Lenticulostriate Arteries (LSA), which supply blood to the basal ganglia~\citep{Fennes2005,Fennes2006}, are the terminal branches of the arterial vascular tree. Pathology of these small vessels is associated with ageing, dementia and Alzheimer's disease~\citep{Wardlaw2013,Zwanenburg2017,chalkias2022differentiating}. Segmentation and quantification of these small vessels is a critical step in the study of Cerebral Small Vessel Disease or CSVD~\citep{duan2020primary,litak2020cerebral}. 
The 7 Tesla (7T) Time-of-Flight MRA is capable of depicting such small vessels non-invasively~\citep{Hendrikse2008,Kang2009}. This paper refers to vessels as small when they have an apparent diameter of only one to two voxels, as seen in 7T MRA with a resolution of 300 \textmu m.   

Manual segmentation of small vessels in 7T MRA is reliable but also time-consuming and laborious. Non-Deep Learning (non-DL) approaches~\citep{frangi1998multiscale,Bernier2018}, which are capable of extracting vessels based on structural properties, can detect medium to large vessels but detecting small vessels can be challenging~\citep{jerman2015beyond}. Manual and semi-automated techniques involve manual efforts in parameter tuning or annotating vessels, making it a time-consuming task requiring experience to achieve the required sensitivity to achieve robust detection of small vessels. Some of the semi-automated techniques use machine learning algorithms to extract features based on the annotations of observers and classify the non-annotated pixels of the image. Another common problem of semi-automatic vessel segmentation~\citep{Sommer2011} is generating spurious points in the segmentation result~\citep{Chen2018}. The aim of this research is to develop an approach that is capable of effectively segmenting small vessels using fully 
automated deep learning techniques based on convolutional neural networks,
 rendering time-consuming parameter tuning or manual intervention obsolete.

In recent years, deep learning-based 
architectures have been widely used for segmentation tasks~\citep{minaee2020image,ulku2020survey}. The 
U-Net~\citep{Ronneberger2015,cciccek20163d} and Attention U-Net~\citep{Oktay2018} have proven to be promising techniques in biomedical image segmentation~\citep{Litjens_2017,liu2021review}. These methods can be trained using small-size datasets and have outperformed the traditional methods~\citep{Ronneberger2015}. Therefore, these architectures are used as baselines for the proposed small vessel segmentation approach. This paper proposes a network combining multi-scale deep supervision~\citep{Zeng2017} in U-Net with elastic deformation consistency learning~\citep{Bortsova2019}. The deep supervision considers the loss at various levels of the U-Net, thereby it should improve the segmentation performance. The consistency learning with elastic deformation makes the model equivariant to these transformations and increases the generalisability. The authors hypothesised that the performance achieved by training this network, which is a combination of these approaches, is better than the existing methods discussed above. Moreover, such a method might also be applicable for segmentation tasks other than blood vessels---while segmenting regions of interest with vastly different sizes and dealing with the problem of the imperfect dataset.

\subsection*{Contribution}
This paper proposes a semi-supervised learning approach, deformation-aware learning DS6, which can learn to perform volumetric segmentation from a small training set. Furthermore, a modified version of U-Net Multi-scale Supervision (U-Net MSS) has been proposed here as the network backbone. The proposed DS6 approach combines elastic deformation with the network backbone using a Siamese architecture. The proposed method has been evaluated for the task of vessel segmentation from 7T ToF-MRA volumes. The proposed method has been compared against previously proposed deep learning and non-deep learning methods for its overall segmentation quality. Moreover, the effect of training set size on the performance of the method and also the performance of the method while segmenting 1.5T and 3T ToF-MRA have been evaluated.  


\section{Related Work}
\label{section:related_work}
In this section, different non-deep learning-based techniques used for vessel segmentation, including semi-automated techniques and deep learning-based techniques, are~discussed.

\subsection{Vessel Segmentation Using Non-DL Techniques}  
The Hessian-based Frangi vesselness filter~\citep{frangi1998multiscale} is one of the most common vessel enhancement approaches and is usually combined with a subsequent, empirically tuned thresholding to obtain the final segmentation. The multi-scale properties of this method enable small vessel segmentation, but considerable parameter fine-tuning might be required to detect the small vessels of interest. 
The authors of
\cite{canero2003vesselness} proposed a vesselness enhancement diffusion (VED) filter, which is a combination of the Frangi filter with an anisotropic diffusion scheme. 
The authors of
\cite{manniesing2005multiscale} extended the VED filter by enforcing the smoothness of the tensor/vessel response function. 
The authors of
\cite{Liao2016} suggested a two-step approach for the segmentation of small vessels. The first step is to detect all the major vessels and most parts of thin vessels using a model-based approach. In the second step, minimal path approaches are used to fill gaps in the initial segmentation and segment-missing low-contrast vessels. However, the minimal path approach involves the manual intervention of observers for key-point detection. Hence, the accuracy of segmentation depends on the information provided by the observer, and this step is time-consuming.
The authors of
\cite{Hsu2017} proposed an automatic vessel filtering pipeline to enhance the contrast of the images. They used the Cardano formula to assign each voxel a value between zero and one as the probability of being a  vessel. This technique requires input filter parameters such as sphericity, tubeness, background suppression, and scale parameter. The output response shows that
 the technique has improved the contrast of small vessels in comparison to the original image and also delineated them well with respect to the background.

The authors of
\cite{Bernier2018} proposed a multi-scale Frangi diffusion filter (MSFDF) pipeline to segment cerebral vessels from susceptibility-weighted imaging (SWI) and TOF‐MRA datasets. The MSFDF pipeline initially pre-selects voxels as vessels or non-vessels by performing a binary classification using a Bayesian Gaussian mixture. This image is further enhanced using the Frangi filter~\citep{frangi1998multiscale} and followed by the VED filter~\citep{manniesing2005multiscale}. Finally, the output is 
intensity-normalised and thresholded to obtain the final segmentation.

The aforementioned methods require manual fine-tuning of the parameters for each dataset, or even for each volume, to obtain the best possible result. Furthermore, these methods require different 
pre-processing techniques, such as bias field correction, making the execution of the pipeline a time-consuming task. 


\subsection{Deep Learning Techniques} 

In recent years, Convolutional Neural Networks (CNN) have outperformed traditional image processing techniques in terms of image classification and image segmentation. These CNNs are extended to more layers to form deep CNNs to achieve further improved performance. This makes the networks huge and requires a larger dataset for training. On the contrary, U-Net~\citep{Raza2017,Blanc-Durand2018,Fabijanska2018,Ronneberger2015} can be fully trained over a smaller set of labelled images. This has made U-Net~\citep{Ronneberger2015,cciccek20163d} a popular architecture in many biomedical image segmentation tasks. 
The authors of 
\cite{Oktay2018} added attention gates to the U-Net architecture to construct attention U-Net, implemented for pancreas segmentation. It was observed that the attention U-Net improves the segmentation accuracy with less computational effort. 
The authors of
\cite{Heller}  showed
 that the U-Net is relatively insensitive to jagged and noisy segmentation boundaries. 
 The authors of
 \cite{girard2019joint} proposed an architecture for joint classification and segmentation of retinal arteries from archive images.   
 The authors of
 \cite{yuan2020hybrid} combined a U-Net with a fully convolutional network to perform vascular segmentation on photo-acoustic imaging. 
 The authors of \cite{yang2019deep} used U-Net for vessel segmentation in X-ray coronary angiography. 
 The authors of
 \cite{livne2019u} used U-Net to perform vessel segmentation on TOF-MRA images of patients with cerebrovascular disease. They used volumes acquired at 3T for training and then validated their approach on images acquired at 7T with a resolution of 600 \textmu m. Motivated by these studies, U-Net was chosen as one of the baseline architectures. 

The authors of
\cite{Zeng2017} proposed an improvement of the U-Net architecture, by computing the overall loss as the sum of losses at each up-sampling scale of the U-Net in its expansion path. This is known as deep supervision~\citep{Zeng2017,Bortsova2017} or multi-scale supervision or MSS~\citep{Zhao2019}. Backpropagating the losses calculated at different scales encourages the architecture to learn discriminative features at every level~\citep{Bortsova2017} and also facilitates more effective gradient flow to earlier stages~\citep{Zeng2017}, which enhances the learning of the network in comparison to U-Net without deep supervision. The authors hypothesise that learning features at different scales will help the model segment vessels of different sizes.
 Hence,
  the authors chose U-Net MSS as the backbone architecture.

The authors of
\cite{Bortsova2019} proposed an architecture for semi-supervised segmentation, where along with supervised learning, the network was made to segment consistently for a given class of transformation. The method was evaluated on a dataset consisting of both labelled and unlabelled chest X-ray images. The model was trained using a supervised loss from the labelled data, and an unsupervised consistency loss was computed between elastically transformed outputs of a Siamese network~\citep{bromley1993signature} made up of U-Nets. Elastic deformation is frequently used in neural networks as a data augmentation technique~\citep{simard2003best} as it can represent variations in the image realistically~\citep{castro2018elastic}, including in the case of vessels. Since this paper dealt with a small dataset using U-Net and further studies have also successfully applied UNet-backed semi-supervised learning   
 for blood vessel segmentation (but in retinal images)~\citep{chen2020semi}, this was considered as a reference and the authors leveraged this strategy to deal with the small dataset at hand.


\section{Proposed Methodology}
\label{section:proposed_methodology}

\begin{figure}[H]
\includegraphics[width=1\textwidth]{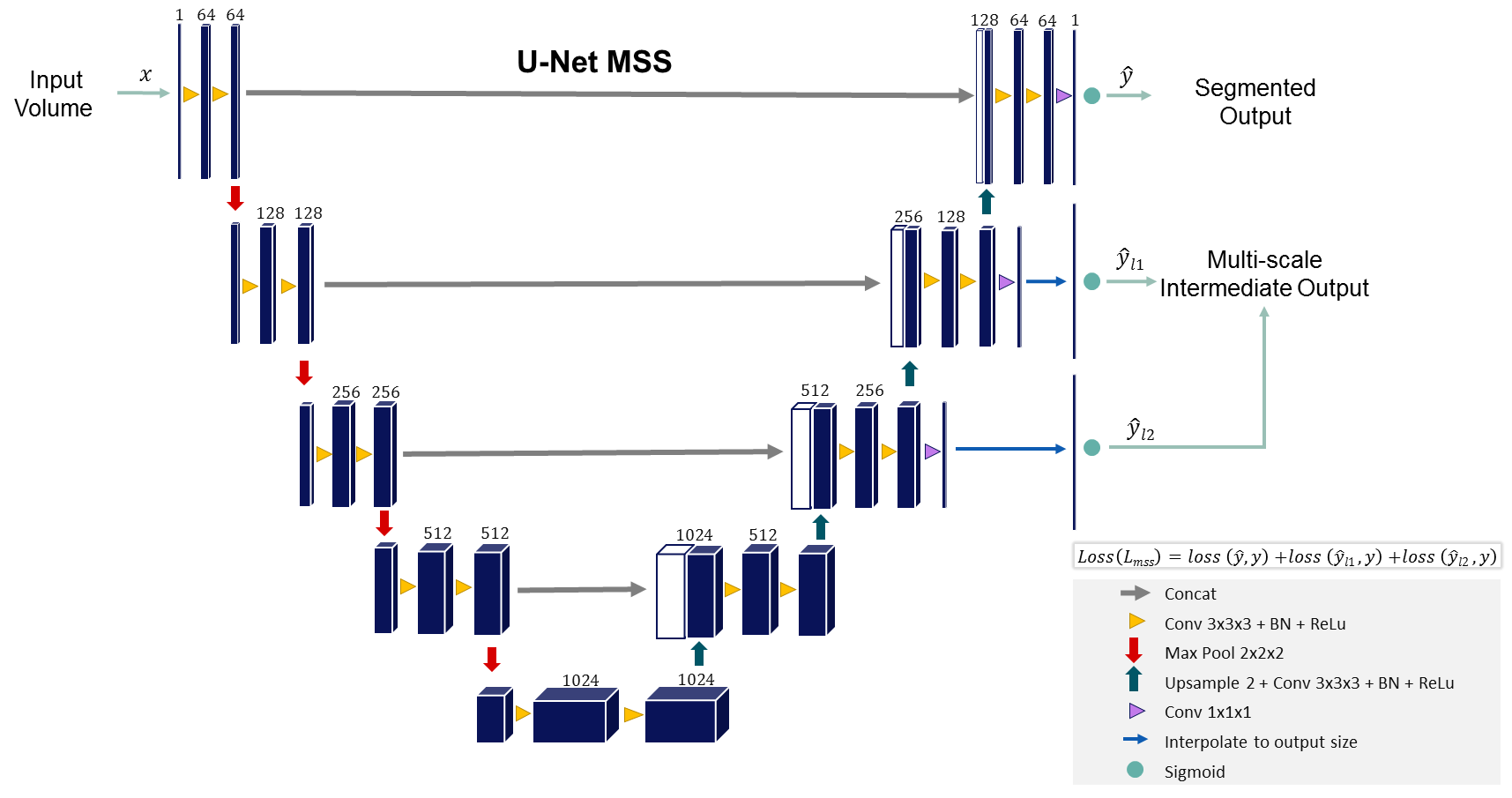}
\caption{Proposed 
 network backbone: Modified U-Net Multi-scale Supervision (Modified U-Net MSS).} \label{fig1}
\end{figure}

The backbone network is a modified version of U-Net MSS~\citep{Zhao2019}. The original U-Net MSS~\citep{Zhao2019} performs five times downsampling using strided convolution and transposed convolution for upsampling, whereas this modified version performs four times downsampling using Max-Pool and performs upsampling using interpolation. Moreover, the original U-Net MSS uses Instance Normalisation and LeakyReLU in its convolution blocks; but this modified U-Net MSS uses Batch Normalisation and ReLU. In addition to the final segmentation output, the outputs of two of the following layers (different scales) are also considered for loss calculation. These additional outputs were interpolated using the nearest neighbour to the size of the segmentation mask before loss calculation. Figure~\ref{fig1} portrays the modified version of the U-Net MSS used in this research. Equation~(\ref{eq:mss_loss}) represents the loss function of the proposed U-Net MSS. The $m$ refers to the total up-sampling scales in the U-Net, \( \ell_{scale}\) is the loss at each up-sampling scale, \( \alpha_i \) is the weight assigned to loss at a specific up-sampling level and network parameter 
 \(\theta\). 
 \vspace{-6pt}
\begin{equation} \label{eq:mss_loss}
\mathcal{L}_{MSS}(\theta) = \frac{1}{\sum\limits_{i=1}^m \alpha_i}\sum\limits_{i=1}^m \alpha_i
\ell_{scale}^i(\theta)
\end{equation}  

\begin{figure}[H]
\includegraphics[width=1\textwidth]{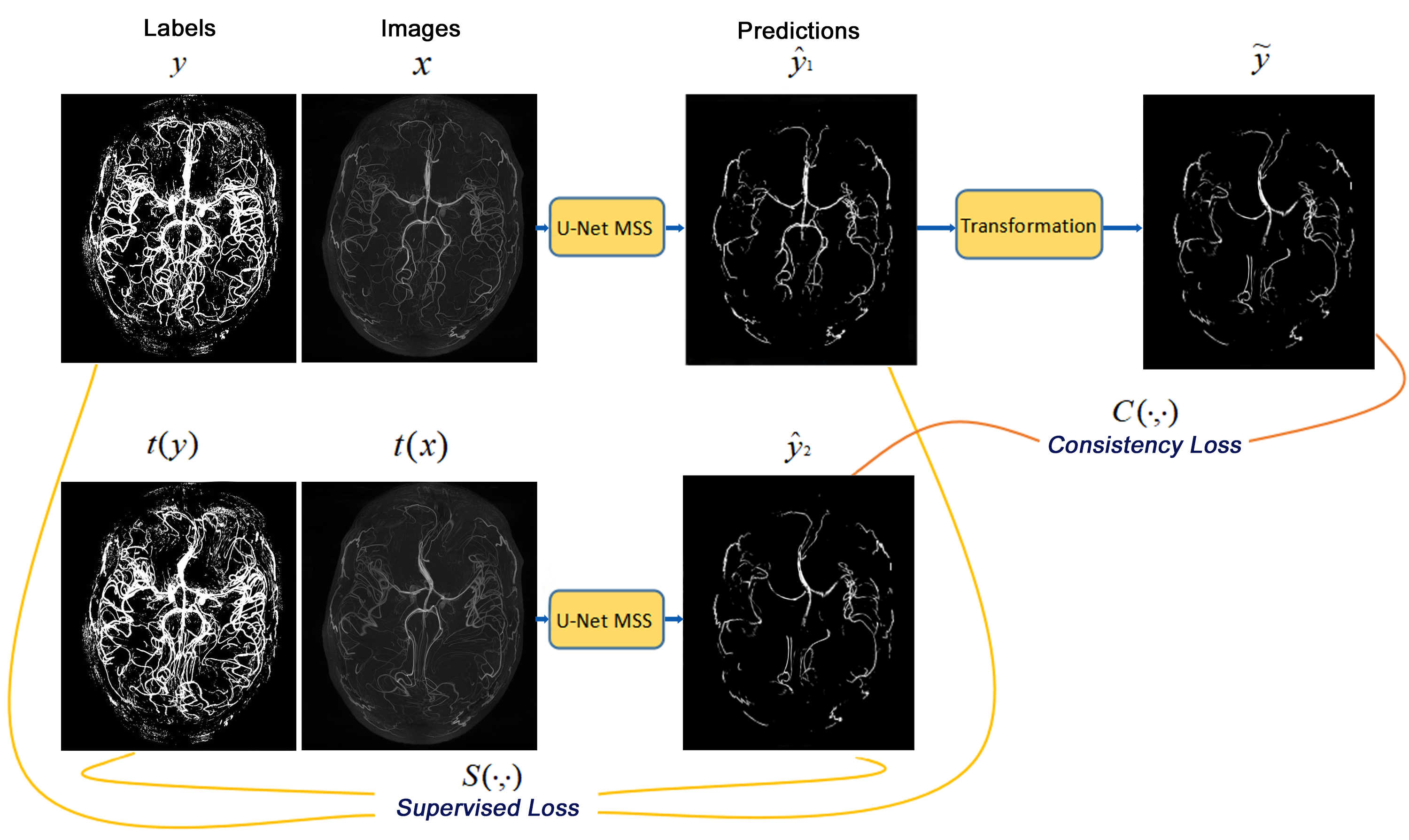}
\caption{The proposed network is based on Siamese architecture, using modified U-Net MSS as the backbone. The original input is fed to the first branch while the second branch receives the elastically deformed version of the input, both branches use the same U-Net MSS for segmentation. The segmentation output of the first branch is elastically deformed using a differentiable transformation layer and is then compared with the output of the second branch for consistency
 (images shown here are only representative and do not portray real results).} \label{fig2}
\end{figure}

To enable the model to learn consistency under elastic deformations~\citep{simard2003best}, the work by~\cite{Bortsova2019} was re-implemented and extended by using the modified U-Net MSS in place of U-Net, as shown in Figure~\ref{fig2}. Let \( \mathcal{X} \) be the set of input volumes while \( \mathcal{Y} \) is the set of corresponding labels and \( \mathcal{T} \) be the set of elastic transformations. The proposed network uses a Siamese architecture to learn from the original data and deformed data using its two identical branches. The first branch is fed with the tuple (\textit{x}, \textit{y}), 
 where \textit{x} \( \sim\mathcal{X} \), \textit{y} \( \sim\mathcal{Y} \), while the second branch is fed with the elastically transformed volume and label (\textit{t}(x),\textit{t}(y)), where \textit{t} \( \sim\mathcal{T} \). These tuples are passed through the U-Net MSS to derive segmentation outputs $\hat{y}_{1}$ and $\hat{y}_{2}$,
  respectively.
   These outputs are compared with the corresponding labels to derive the supervised loss depicted in Equation~(\ref{eq:supervised_loss}). Furthermore, the $\hat{y}_{1}$ is elastically transformed by \textit{t} 
 $\hat{y}_{1}$ to \( \Bar{y} \).
  Now,
   the \( \Bar{y} \) is compared with $\hat{y}_{2}$ for computing the consistency loss in self-supervised manner as shown in Equation~(\ref{eq:consistency_loss}). The network was trained to find the optimal value of \(\theta\) that minimises the overall loss defined by the sum of the supervised and the consistency loss.
 \vspace{-6pt}
\begin{equation} \label{eq:supervised_loss}
\mathcal{L}_{Sup}^\tau(\theta) =  \frac{1}{\mid\mathcal{X}\mid}  \sum\limits_{x\in\mathcal{X}, t \in \tau} \mathcal{L}_{MSS}(x,y,\theta) + \mathcal{L}_{MSS}(t(x),t(y),\theta) 
\end{equation}
\begin{equation} \label{eq:consistency_loss}
\mathcal{L}_{Cons}^\tau(\theta) = \frac{1}{\mid\mathcal{X}\mid}  \sum\limits_{x\in\mathcal{X}} \mathbb{E}_{t \sim \tau}[\mathcal{C}(t(f(x;\theta), f(t(x);\theta))]
\end{equation}

The authors hypothesise that this proposed network can learn consistency under elastic transformations and will be able to learn from the available small dataset with noisy labels. The supervised loss \(\mathcal{L}_{Sup}^\tau(\theta)\) trains the network to segment the images using the ground-truth labels, and the consistency loss \(\mathcal{L}_{Cons}^\tau(\theta)\) ensures that the network is equivariant to the elastic transformation. This transformation can also be used as data augmentation for the models, but this was not applied in this research work as it was already shown by \cite{Bortsova2019} that the deformation consistency performs better than using it as data augmentation. 

\vspace{-15pt}

Elastic deformation of a given volume can be obtained by applying a displacement field to the volume, as shown in Equation~(\ref{eq:elastic}), where $\Delta x_{x, y, z}$, $\Delta y_{x, y, z}$ and $\Delta z_{x, y, z}$ are the displacements in x, y and z 
 direction, respectively;
  g denotes the grey values of each voxel of the original image, whereas $\mathrm{g}^{\prime}$ denotes the grey values of the deformed image~\citep{song2016handwritten}. The displacement fields were randomly generated by initialising a grid with random values within a specified limit, followed by interpolating the displacement at each voxel from the grid using 
  a 
  cubic B-spline~\citep{knott2000interpolating}.
 \vspace{-6pt} 
\begin{equation} \label{eq:elastic}
\mathrm{g}^{\prime}(x, y, z)=\mathrm{g}\left(x+\Delta x_{x, y, z}, y+\Delta y_{x, y, z},z+\Delta z_{x, y, z}\right)
\end{equation}


\section{Datasets and Labels}
\label{sec:DS}
To be able to validate the proposed approach, a high-resolution TOF-MRA dataset was needed. Images obtained with high spatial resolution on a 7T MR scanner contain more small vessels in comparison to a 3T MR scanner~\citep{Liao2016}. Hence, the dataset by~\cite{Mattern2018} was chosen, which comprises 
 3D TOF-MRA of 11 subjects imaged at 7T. 
 The data had an isotropic resolution of 300 \textmu m and were acquired with prospective motion correction to prevent image blurring, hence preventing the loss of small vessels~\citep{Mattern2018,sciarra2022quantitative}.

For this dataset, no ground-truth segmentation was available. The labels for these MRA images were created semi-automatically using Ilastik~\citep{Ilastik} by a computer scientist (see Acknowledgement), who was initially trained, and later,
 the segmentation quality was checked visually by a domain expert (H.M.). 
  Ilastik is a widely used tool for pixel classification, which provides the capability to create custom pixel classification workflows. The annotator can visualise the volume and then
  annotate the small vessels using points or scribbles by adjusting the pointer diameter. Ilastik uses these annotations and pixel features of the image to segment small vessels in the rest of the volume using the Random Forest algorithm. Further, the user can go through the rest of the slices and manually modify the segmentation. Alternatively, Ilastik can also be used for fully manual segmentation. For the label creation task on the dataset, Ilastik's semi-automatic option with manual interaction was used.

\begin{figure}[H]
\includegraphics[width=0.46\textwidth]{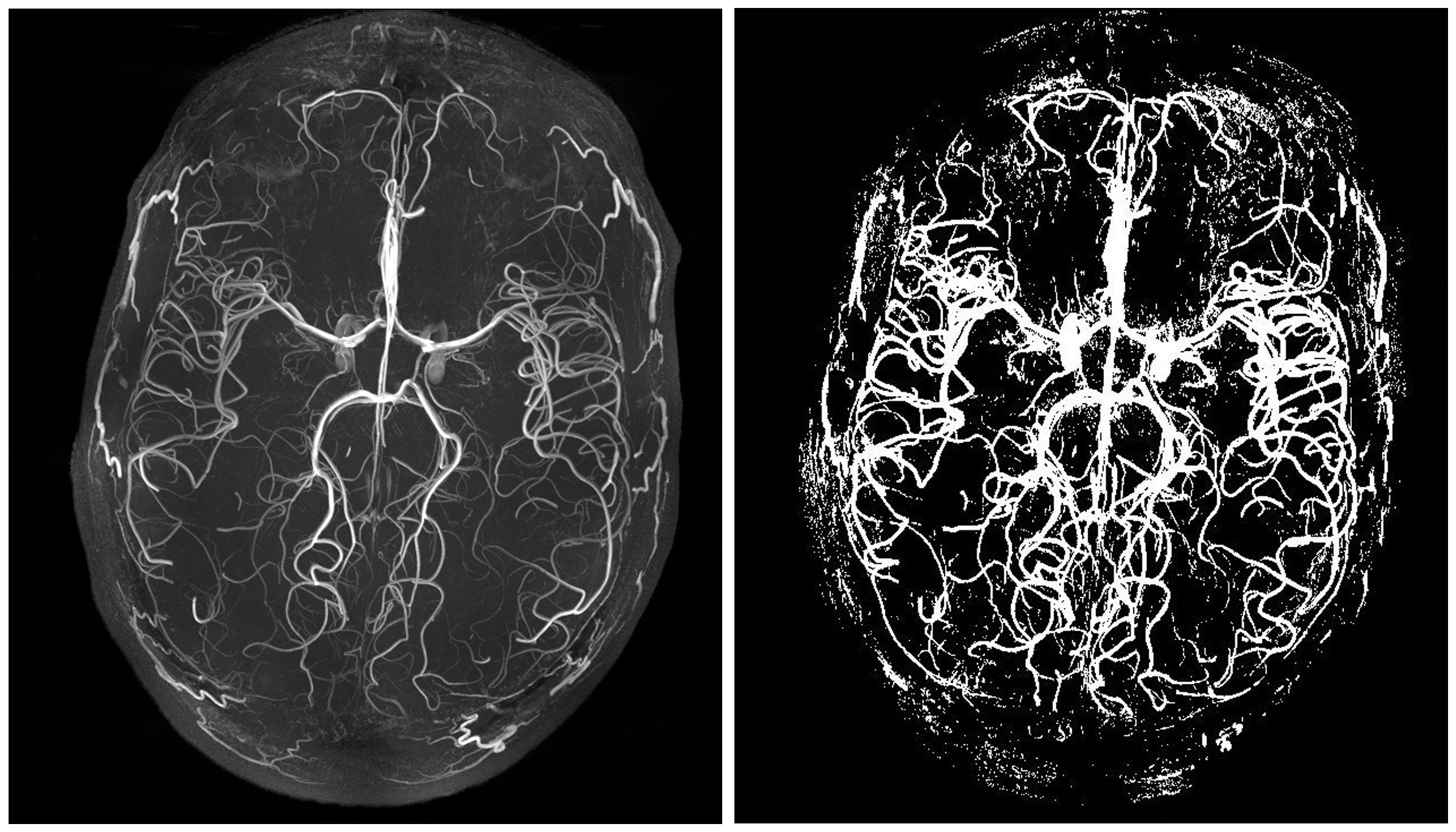}
\caption{Maximum intensity projection (MIP) of the actual volume and the segmentation labels created using Ilastik.}  
\label{fig3}
\end{figure}

The quality of the labelled images can be visually assessed by comparing the maximum intensity projection (MIP) of the original volume with the label, which is shown in Figure~\ref{fig3}. The label generated is not entirely perfect, as it is noisy, contains gaps in certain regions, and does not capture all small vessels, which can be attributed to the limitations of the tool. For a more accurate analysis of the method, the fully manual segmentation option of Ilastik was used to manually segment a small region of interest (ROI) of dimension 64$^{3}$, as can be seen in 
 Figure~\ref{fig7}.

For training and evaluation of the network, the dataset was randomly divided into training, validation, and test sets in the ratio of 6:2:3. The performance of the methods was evaluated using 3-fold cross-validation.  
For further evaluation of the generalisation capabilities of the proposed network, 
 publicly available MRA images 
 from the IXI dataset (IXI dataset: 
 \url{https://brain-development.org/ixi-dataset/ (Accessed on: 21.09.2022)}) were used. The data were collected on two different field strengths, 1.5T and 3T. The labels for these images were created using Ilastik using a semi-automated manner, as mentioned earlier.


\section{Experimental Setup}
The Siamese network used in this research consists of a U-Net MSS~\citep{Zeng2017} as the backbone architecture, as explained in the previous section. 
Every MRA image volume was converted to 3D patches with dimensions 64$^{3}$, and with a stride of 32, 32, 16 
 across sagittal (width), coronal (height), and axial (depth), respectively, to get overlapped samples---to allow the network to learn inter-patch continuity. Having six image volumes for training and two volumes for validation, this configuration created 20,412 patches for training and 6804 for validation. For additional trainings to judge the effect of the training set size on the models, three additional training sets with one, two, and four volumes were also created. For the training of the deep learning models, 8000 patches were selected at random in each epoch and fixed 6000 patches for validation---to reduce the number of iterations per epoch, reducing the required training time. For evaluation, the three test volumes were converted to non-overlapping patches of size 64$^{3}$ using strides of 64 across all dimensions, this resulted in 1012 patches.
For the calculation of the supervised loss and the consistency loss, the focal Tversky loss or FTL~\citep{focalT} as in Equation~(\ref{eq:focal_traversky_loss}) was used and was optimised during training using the Adam optimiser~\citep{kingma2014adam} with a learning rate of 0.01. The range of $\gamma$ is between 1 and 3. The Tversky Index (TI) was calculated using 
 Equation~(\ref{eq:traversky_loss}), where $\alpha$ and $\beta$ are the parameters to decide the weight given to the false negatives and to the false positives. The \({p_{ic}}\) represents the probability of voxel i being vessel, and \({p_{i\bar{c}}}\) represents the probability of pixel i being non-vessel. Similarly, the \({g_{ic}}\) and \({g_{i\bar{c}}}\) represent the label of the pixel with value either 0 or 1. In this research, $\alpha$ was chosen to be 0.7 and $\beta$ to be 0.75 after~experimentation. 
\begin{equation} 
\label{eq:focal_traversky_loss}
\mathcal{FTL}_{c} = (\sum\limits_{c} 1-\mathcal{TI}_{c})^\frac{1}{\gamma}
\end{equation}
\begin{equation} 
\label{eq:traversky_loss}
\mathcal{TI}_{c} = \frac{\sum\limits_{i=1}^N p_{ic}g_{ic} + \epsilon }{\sum\limits_{i=1}^N p_{ic}g_{ic} + \alpha \sum\limits_{i=1}^N p_{i\bar{c}}g_{ic} + \beta \sum\limits_{i=1}^N p_{ic}g_{i\bar{c}} + \epsilon }
\end{equation}

The implementation of the deep learning models was done using PyTorch~\citep{paszke2019pytorch}. A differentiable version of the elastic transformation was implemented by extending the random elastic transformation of TorchIO~\citep{fern2020torchio} by replacing its SimpleITK-based~\citep{lowekamp2013design} implementation of the transformation with a modified version of the kernel transformation (using the 3D B-spline kernel) of AirLab~\citep{Airlab}. The number of control points along each dimension of the grid (see Section
~\ref{section:proposed_methodology}) was chosen randomly to be five, six or seven, to introduce different levels of deformation. The maximum displacement was set to 0.02, and two~borders were locked. Elastic transformations, along with the selection of its number of control points, were performed randomly for each patch for each training iteration. In this manner, the network encountered differently deformed patches with different levels of deformations during the training process. The code of this implementation can be obtained from GitHub (
 \url{https://github.com/soumickmj/DS6} - Accessed on 21.09.2022).

As all the models (proposed method and the baselines) converged between 20 and 30~epochs, the trainings were performed for 50 epochs with mixed precision~\citep{micikevicius2017mixed} using Nvidia Apex 
 (\url{https://github.com/NVIDIA/apex} - Accessed on 8 March 2020, commit 5633f6d) on Nvidia Tesla V100-SXM2-32GB GPU. Using Mixed precision, it was possible to train with a batch size of 20, in comparison to a batch size of eight with Full precision (more details in Appendix~\ref{app:mixedprecis}). 
 The lowest validation loss was noted, and the corresponding epoch's model weights were saved. These trained models were then made deformation-aware by further training for 50 more epochs with the proposed Siamese architecture.


\section{Evaluation}
\label{section:evaluation}
In this section, the performance of the proposed method is compared against that of other methods. Two different non-deep learning (Non-DL)-based methods were used for comparison---Frangi filter~\citep{frangi1998multiscale}, which is one of the most widely used methods for vessel segmentation and the MSFDF pipeline~\citep{Bernier2018}, which is one of the state-of-the-art methods for vessel segmentation for SWI and TOF image volumes. For both of these methods, the volumes were pre-processed  
 with N4 bias field correction~\citep{tustison2010n4itk} to remove intensity inhomogeneities, which is a common effect in high-field MR images resulting in spatial intensity variations~\citep{truong2006effects}. Multiple experiments were performed with manual tunings of the parameters of the Frangi filter to obtain a suitable parameter set. Finally, a three-scale (1, 2, 3) 
 Frangi filter with a $\gamma$ of 0.1 (Frangi correction constant) and a final threshold of 0.01 was chosen for this dataset and was applied on pre-normalised images (between 0--1). For the MSFDF pipeline, the official implementation 
 (\url{https://github.com/braincharter/vasculature_notebook} - Accessed on 10 December 2020, commit 0bb7244) was used with all the default parameters as provided, except for the Otsu offset, which was set to 0.0 (default 0.2) to reduce noise contamination in the final segmentation. Besides the bias-field corrected image volumes, the method required a brain mask. To that end, the brain extraction tool or BET~\citep{smith2002fast,jenkinson2005bet2} of the FSL~\citep{jenkinson2012fsl} was used (fractional threshold set to 0.1). 

The performance of the proposed model was compared against two established deep learning-based baseline models---U-Net~\citep{cciccek20163d}, for being one of the most popular methods for segmentation and Attention U-Net~\citep{Oktay2018}, which was proposed as an upgrade of U-Net. The performance of the proposed method was compared with and without the deformation consistency; the Siamese network U-Net MSS + Deformation (proposed method) and the modified U-Net MSS (proposed backbone), respectively. The proposed method was also compared against a re-implemented version U-Net + Deformation~\citep{Bortsova2019}. This re-implemented version had exactly the same setup as the proposed method, except for the network model. No pre-processing, such as bias field correction or brain extraction, was performed on the images before supplying them to the deep learning models, except for normalising the intensities (between 0--1).

Quantitative evaluation of the 3D segmentations was performed using the Dice coefficient and the Intersection over Union (IoU), also known as the Jaccard Index. For qualitative analysis, maximum intensity projections (MIP) were computed to provide an overview of the 3D data as a 2D representation. MIPs were computed for whole brain assessment as well as ROI-specific. ROI-specific MIPs enable visual comparison of the small vessel segmentation quality of the different approaches used.
 
The first set of evaluations was performed by comparing the results against the 7T MRA dataset labels, which were created semi-automatically with Ilastik (see Section~\ref{sec:DS}). Then,
 the effect of the training set size on the results was evaluated. Moreover, given that the dataset labels are noisy, evaluations of the models were performed by comparing the results against a manually segmented ground-truth. Lastly, to evaluate the generalisation capabilities of the models, they were employed to segment 1.5T and 3T MRA volumes.


\subsection{On 7T MRA Test Set}

The first set of evaluations was performed for three out of eleven volumes of the 7T~MRA dataset with 300 \textmu m isotropic resolution (test set separated from the dataset before training/validation).

As shown in Table~\ref{tab1}, the Frangi filter and the method proposed in~\cite{Bernier2018}  
 did not perform competitively with Deep Learning methods quantitatively. Further, qualitatively,
 it is evident from Figure~\ref{fig4} that both Non-Deep Learning methods failed to detect small vessels in most of the cases. The Frangi filter failed to detect small vessels in columns three and four.
  In the method proposed in \cite{Bernier2018}, 
  non-vessel regions were
   segmented as vessels, which is evident from the dominant blue in columns one and two. All deep neural networks outperformed traditional techniques by a large margin (51.81 $\pm$ 3.09 for Frangi: the best non-DL baseline, 
    76.19 $\pm$ 0.17 for baseline U-Net: the worst DL baseline) even with imperfectly labelled data. Careful visual inspection of the MIPs showed that U-Net MSS segmented the vessels better than the rest of the models. Further, on applying deformation-aware learning, a considerable improvement (4.27\%) was observed with U-Net and also with U-Net MSS (1.37\%). This can also be further noticed in the 
  evaluation with different training set sizes conducted on the same models, where deformation-aware learning improved the performance of the~models.

\begin{table}[H]
\caption{Comparison of metrics for Non-Deep Learning and Deep Learning methods.}
\label{tab1}
\newcolumntype{C}{>{\centering\arraybackslash}X}
\begin{tabularx}{\textwidth}{m{2cm}<{\centering}m{5cm}<{\centering}CC}
\toprule
\textbf{Type} & \textbf{Method} &  \textbf{Dice Coeff.} & \textbf{IoU}\\
\midrule
Non-DL & Frangi Filter &  {51.81 \(\pm\) 3.09} & {35.00 \(\pm\) 2.85}\\
\midrule
Non-DL & MSFDF Pipeline &  {48.35 \(\pm\) 6.34} & {32.04 \(\pm\) 5.55}\\
\midrule
DL & Attention U-Net & {76.73 \(\pm\) 0.22} & {62.25 \(\pm\) 0.29}\\
\midrule
DL & U-Net &  {76.19 \(\pm\) 0.17} & {61.54 \(\pm\) 0.22}\\
\midrule
DL & \textbf{U-Net MSS 
} & {79.35 \(\pm\) 0.35} & {65.81 \(\pm\) 0.47}\\
\midrule
DL & \textbf{U-Net + Deformation} & {79.44 \(\pm\) 0.89} & {65.97 \(\pm\) 1.23}\\
\midrule
DL & \textbf{U-Net MSS + Deformation} & {80.44 \(\pm\) 0.83} & {67.37 \(\pm\) 1.16}\\
\bottomrule
\end{tabularx}
\end{table}

\subsubsection*{Statistical Hypothesis Testing}
The statistical significance of the improvements observed was analysed by means of an independent two-sample \textit{t}-test. As the test set of this research contains only three volumes and the population size might be too small to perform a reliable hypothesis testing, an additional set of Dice scores was calculated by dividing the volumes on the x-axis into nine equal non-overlapping (independent) slabs---resulting in slabs with dimensions 80 $\times$ 630 \linebreak $\times$ 195. It was observed that deformation-aware learning resulted in significant improvement for both U-Net and U-Net MSS (\textit{p}-values 0.004 and 0.013, respectively). Furthermore, the incorporation of multi-scale supervision with U-Net resulted in a significant improvement (\textit{p}-value 0.003). However, U-Net MSS with deformation did not show any significant improvement over U-Net with deformation (\textit{p}-value 0.086).


\subsection{On the Effect of the Training Set Size}

Further experiments were performed with varying training set sizes to understand its influence on the performance of the models and to understand how a lower number of volumes can be used for training the models to archive adequate performance. Four different training set sizes were experimented with: Six, Four, Two, and One. The sizes of the validation and test set were kept fixed (two and three, respectively). A 3-fold cross-validation was used for each training set size.

Table~\ref{tab2} shows the results obtained by training the networks with different training set sizes, and Figure~\ref{fig5} portrays the resultant Dice scores using violin plots to observe the data distribution. From the violin plots, it can be observed that the median values (the middle-dotted lines inside the violins) are higher in every case for deformation-aware models. Moreover, it can be observed that the models trained with six volumes performed better than the models trained with smaller training sets and the models trained with one volume performed the worst. The result of the models trained with four and two volumes are very similar, but the results of trainings with two volumes are better than four. Given that the dataset labels are imperfect, it could have contributed to these counter-intuitive results, but this requires further investigation. 

\begin{figure}[H]
\includegraphics[width=0.75\textwidth]{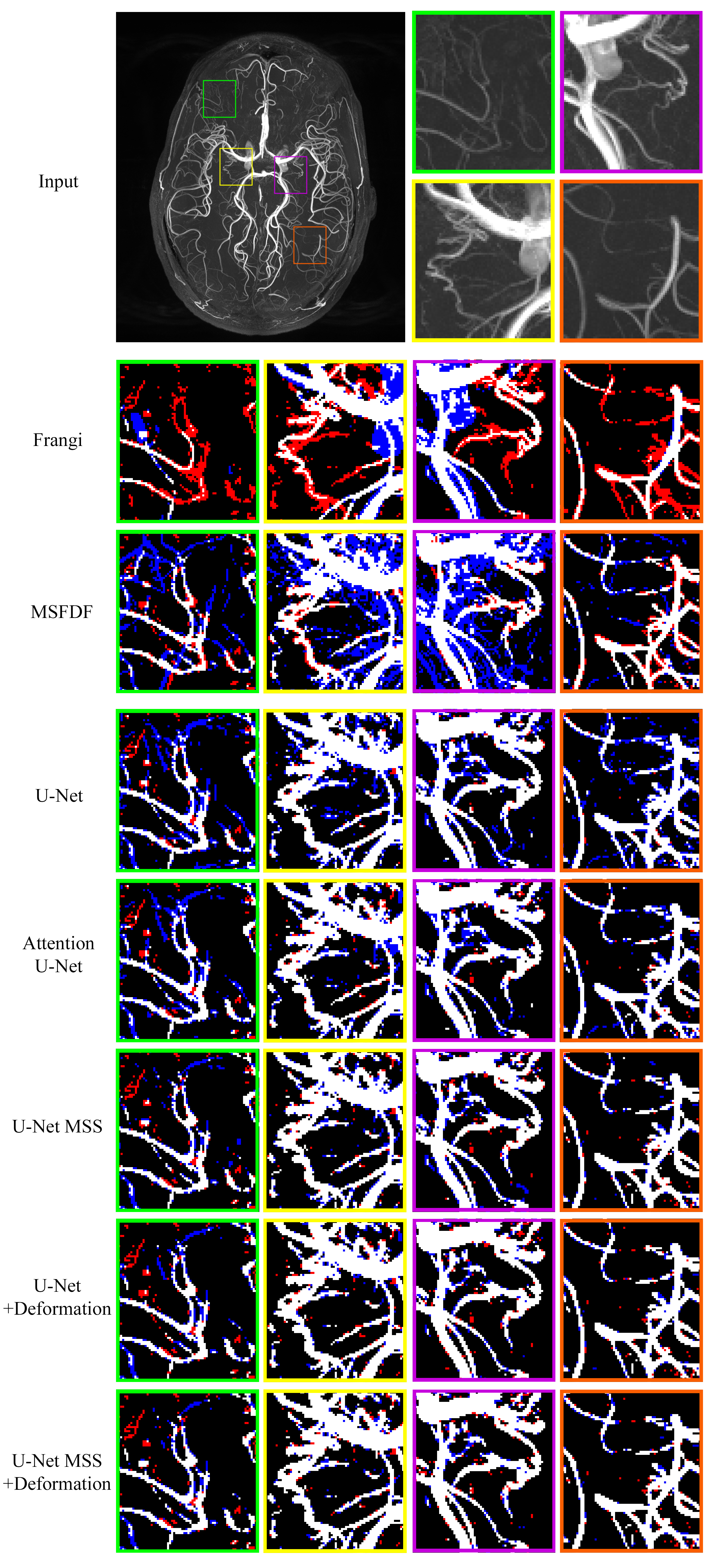}
\caption{Comparisons of different non-DL and DL methods with the maximum intensity project (MIP) of the input volume. Red indicates false negative and blue indicates false positive (MSFDF: multi-scale Frangi diffusion filter, U-Net MSS: U-Net multi-scale supervision)}.        
\label{fig4}
\end{figure}

\vspace{-6pt}
\begin{figure}[H]
    \includegraphics[width=\textwidth]{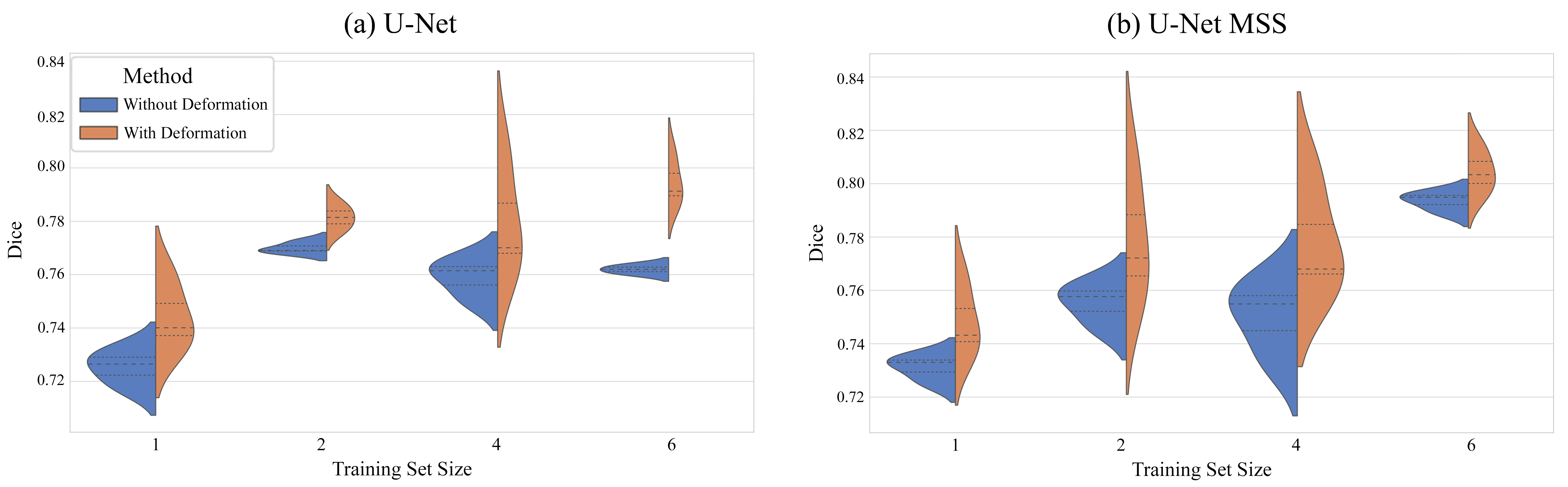}
\caption{Violin 
 plots showing the quantitative results of the experiments with training set sizes, for U-Net (\textbf{a}) and U-Net MSS (\textbf{b}) trained with and without deformation-aware learning.}  
\label{fig5}
\end{figure}
\vspace{-10pt}

\begin{table}[H]
\caption{Quantitative evaluation of the effect of training set sizes on U-net and U-net MSS with and without deformation-aware learning.}
\label{tab2}
\begin{adjustwidth}{-\extralength}{0cm}
		\newcolumntype{C}{>{\centering\arraybackslash}X}
		\begin{tabularx}{\fulllength}{CCCCCC}
\toprule
\multicolumn{2}{c}{\textbf{}} &
  \multicolumn{2}{c}{\textbf{Without Deformation}} &
  \multicolumn{2}{c}{\textbf{With Deformation-Aware  Learning}} \\ 
  \midrule
\textbf{Model} &
  \textbf{Training Set Size} &
  \textbf{Dice Coeff.} &
  \textbf{IoU} &
  \textbf{Dice Coeff.} &
  \textbf{IoU} \\ 
  \midrule
\multirow{4}{*}{U-Net 
} &
  1 &
  72.52 \(\pm\) 0.67 &
  56.91 \(\pm\) 0.81 &
  74.40 \(\pm\) 1.25 &
  59.25 \(\pm\) 1.60 \\
 &
  2 &
  76.99 \(\pm\) 0.21 &
  62.73 \(\pm\) 0.36 &
  78.13 \(\pm\) 0.47 &
  64.28 \(\pm\) 0.61 \\
 &
  4 &
  75.88 \(\pm\) 0.72 &
  61.19 \(\pm\) 0.89 &
  77.97 \(\pm\) 2.06 &
  63.96 \(\pm\) 2.82 \\
 &
  6 &
  76.19 \(\pm\) 0.17 &
  61.54 \(\pm\) 0.22 &
  79.44 \(\pm\) 0.89 &
  65.97 \(\pm\) 1.23 \\ 
  \midrule
\multirow{4}{*}{\textbf{U-Net MSS}} &
  1 &
  73.11 \(\pm\) 0.48 &
  57.63 \(\pm\) 0.57 &
  74.81 \(\pm\) 1.32 &
  59.78 \(\pm\) 1.70 \\
 &
  2 &
  75.52 \(\pm\) 0.78 &
  60.76 \(\pm\) 1.07 &
  77.84 \(\pm\) 2.35 &
  63.95 \(\pm\) 3.19 \\
 &
  4 &
  75.02 \(\pm\) 1.36 &
  60.10 \(\pm\) 1.65 &
  77.79 \(\pm\) 2.05 &
  63.76 \(\pm\) 2.88 \\
 &
  6 &
  79.35 \(\pm\) 0.35 &
  65.81 \(\pm\) 0.47 &
  80.44 \(\pm\) 0.83 &
  67.37 \(\pm\) 1.16 \\ 
  \bottomrule
		\end{tabularx}
	\end{adjustwidth}
\end{table}

\begin{figure}[H]
\includegraphics[width=0.49\textwidth]{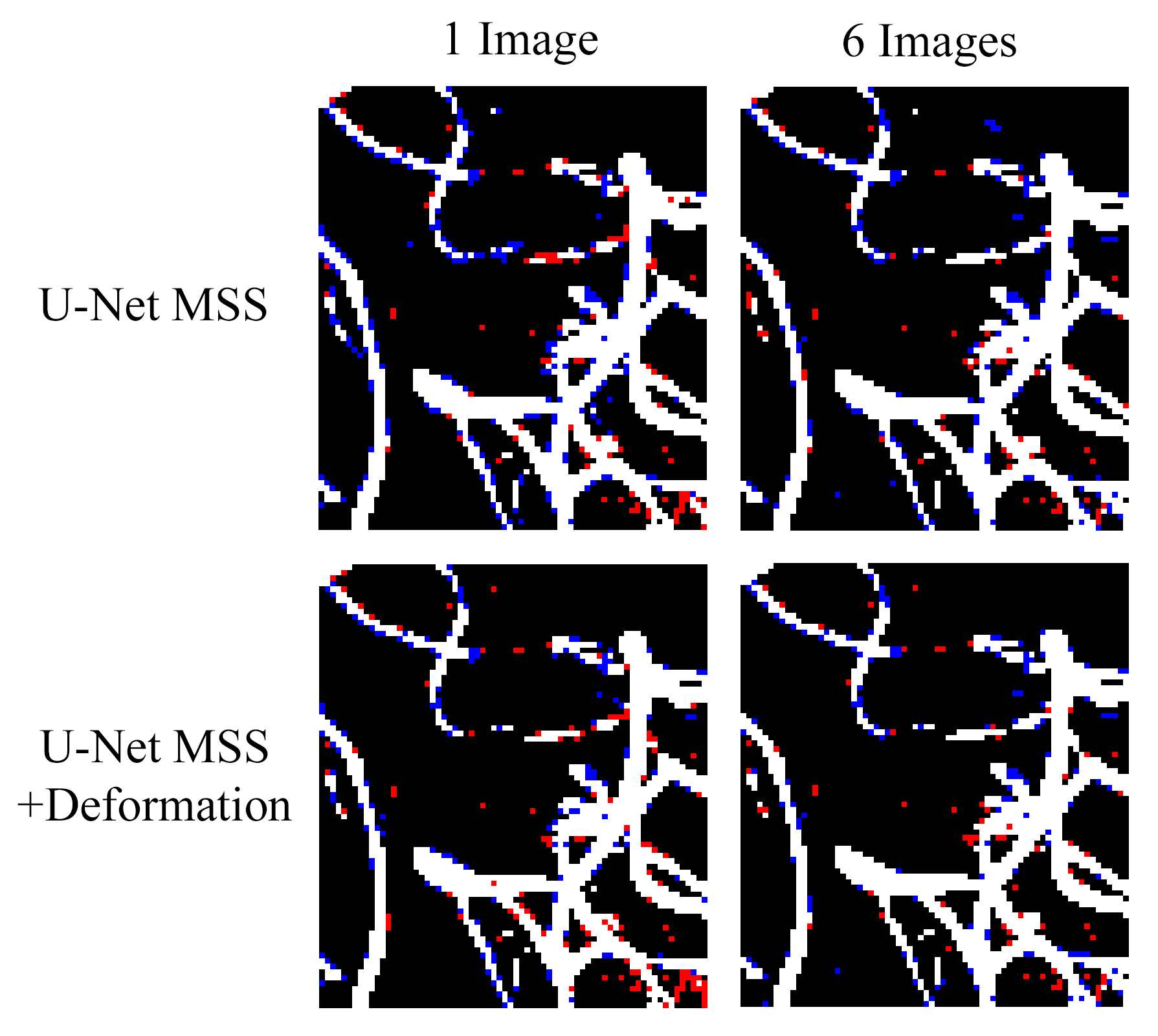}
\caption{Qualitative comparison of models trained with one volume and six volumes, for U-Net MSS with and without deformation-aware learning. Red indicates false negative and blue indicates false positive.}
\label{fig6}
\end{figure}

\vspace{-6pt}
Models trained on a single volume still show good segmentation of vessels \linebreak (74.81 $\pm$ 1.32 Dice for  U-Net MSS + Deformation), which are comparable with the ones trained with six volumes even though typical deep learning models expect to have a large dataset. It can be further observed in Figure~\ref{fig6} that U-Net MSS trained with deformation-aware learning could maintain certain vessel continuity better than its counterpart, which was trained without deformation.

\subsection{On a Manually Segmented ROI of 7T MRA} 
The labels created semi-automatically using Ilastik were imperfect (as discussed in Section~\ref{sec:DS}). Hence, a single 64$^{3}$ ROI was segmented manually from an input image of the 7T~MRA test set, and the performance of the methods was compared against this manually segmented ROI (considered perfect label, hence, ground truth).

\begin{figure}[H]
\includegraphics[width=\textwidth]{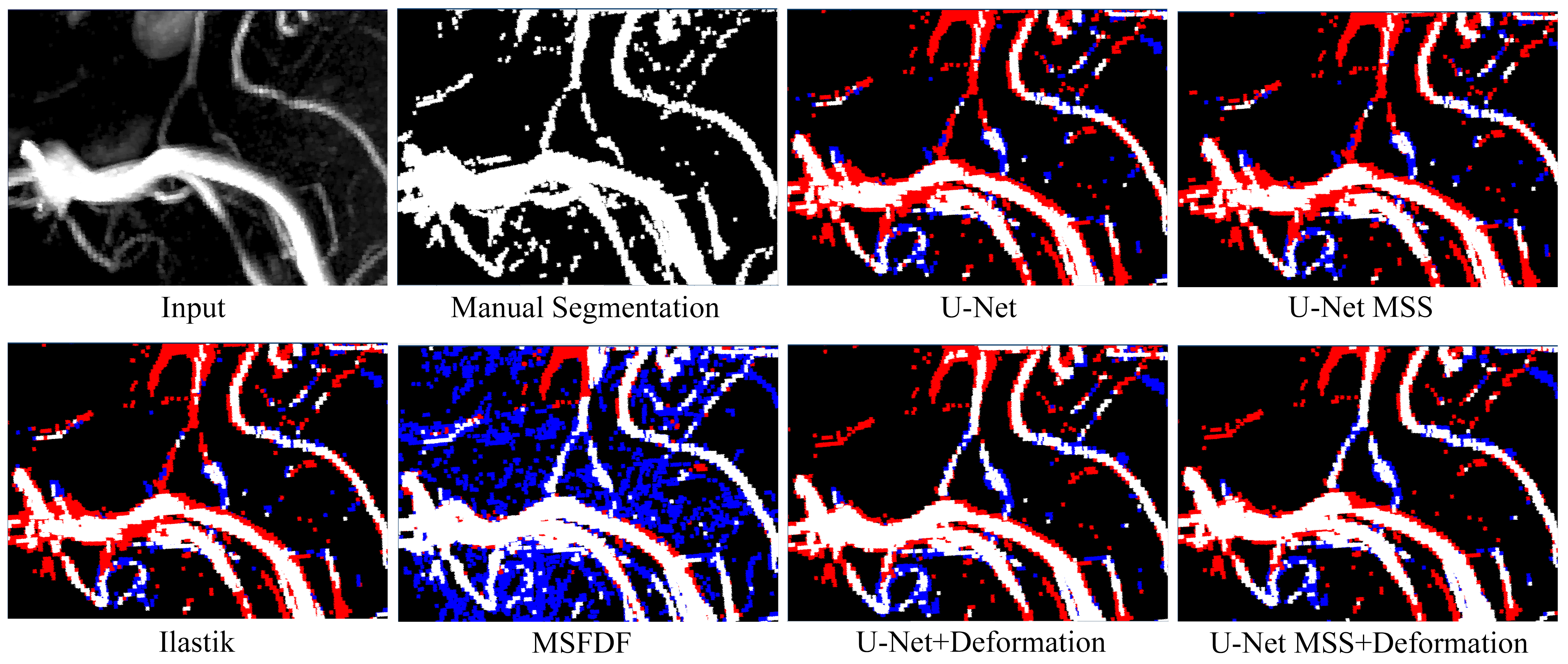}
\caption{Qualitative evaluation of various methods with ROI-specific MIPs. The ROI was selected to feature large to small-scale vessels and enable qualitative comparison across segmentation methods. Red colour indicates the false negative while blue represents the false positive while being compared against manual segmentation.}        
\label{fig7}
\end{figure}

Figure~\ref{fig7} shows a qualitative analysis comparing the manual segmentation against the prediction from all methods. The comparison shows that the semi-automated segmentation with Ilastik and all the baselines failed to detect the entire Y-shaped small vessels at the centre of the ROI. On the contrary, the models trained with deformation were able to detect these Y-shaped small vessels and maintain vessel continuity. Further, for quantitative analysis, the Dice scores of the predictions against the manually segmented ROI were calculated (see Table~\ref{tab3}). It can be observed that the training labels created with Ilastik achieved only a 50.21 Dice while being compared against the manually segmented ROI. Furthermore, U-Net MSS resulted in 9.95\% better Dice than U-Net, and deformation-aware U-Net MSS resulted in
 18.98\% better Dice than the U-Net MSS without deformation.

\vspace{-6pt}

\vspace{-10pt}

\begin{table}[H]
\caption{Resultant Dice coefficients while comparing different methods against a manually segmented~region.}
\label{tab3}
\newcolumntype{C}{>{\centering\arraybackslash}X}
\begin{tabularx}{\textwidth}{CCC}
\toprule
\textbf{Type} & \textbf{Method} &  \textbf{Dice Coeff.} \\

\midrule
non-DL & MSFDF &  {52.39} \\
\midrule
non-DL & Frangi &  {57.59} \\

\midrule
Training Labels & Ilastik &  {50.21} \\

\midrule
DL & U-Net &  {47.45} \\
\midrule
DL & \textbf{U-Net MSS 
} &  {52.17} \\
\midrule
DL & \textbf{U-Net + Deformation} &  {59.81} \\
\midrule
DL & \textbf{U-Net MSS + Deformation} &  {62.07} \\
\bottomrule
\end{tabularx}
\end{table}

\vspace{-10pt}

\subsection{On Publicly Available Dataset
 with Lower Resolution than the Training Set: IXI MRA}  

To evaluate the generalisation capabilities of the model, it was applied to the publicly available IXI MRA dataset. The data were acquired at 1.5T and 3T with a lower image resolution of 450 \textmu m. Due to this reduced resolution, the IXI dataset is less sensitive to small vessels (based on the true vessel lumen and not the apparent vessel diameter in voxel). The resolution of this dataset (450 \textmu m) is 1.5 times lower in each direction in comparison to the training dataset (300 \textmu m) acquired using a 7T scanner. This difference in resolution motivated the authors to perform experiments with two additional patch sizes (32$^{3}$ and 96$^{3}$) during inference apart from 64$^{3}$, which has been used during training. It can be observed that the patch size 96$^{3}$ gave the best performance, which is 1.5 times higher than the training patch size. 

Table~\ref{tab4} summarises the quantitative performance of the models on 1.5T and 3T images. It shows that, in general, the U-Net MSS model with deformation consistency outperforms the model without deformation consistency on both image sets (10.84\% and 3.42\% improvement with deformation for 1.5T and 3T, respectively, for patch size 96$^{3}$).

\begin{figure}[H] 
\includegraphics[width=\textwidth]{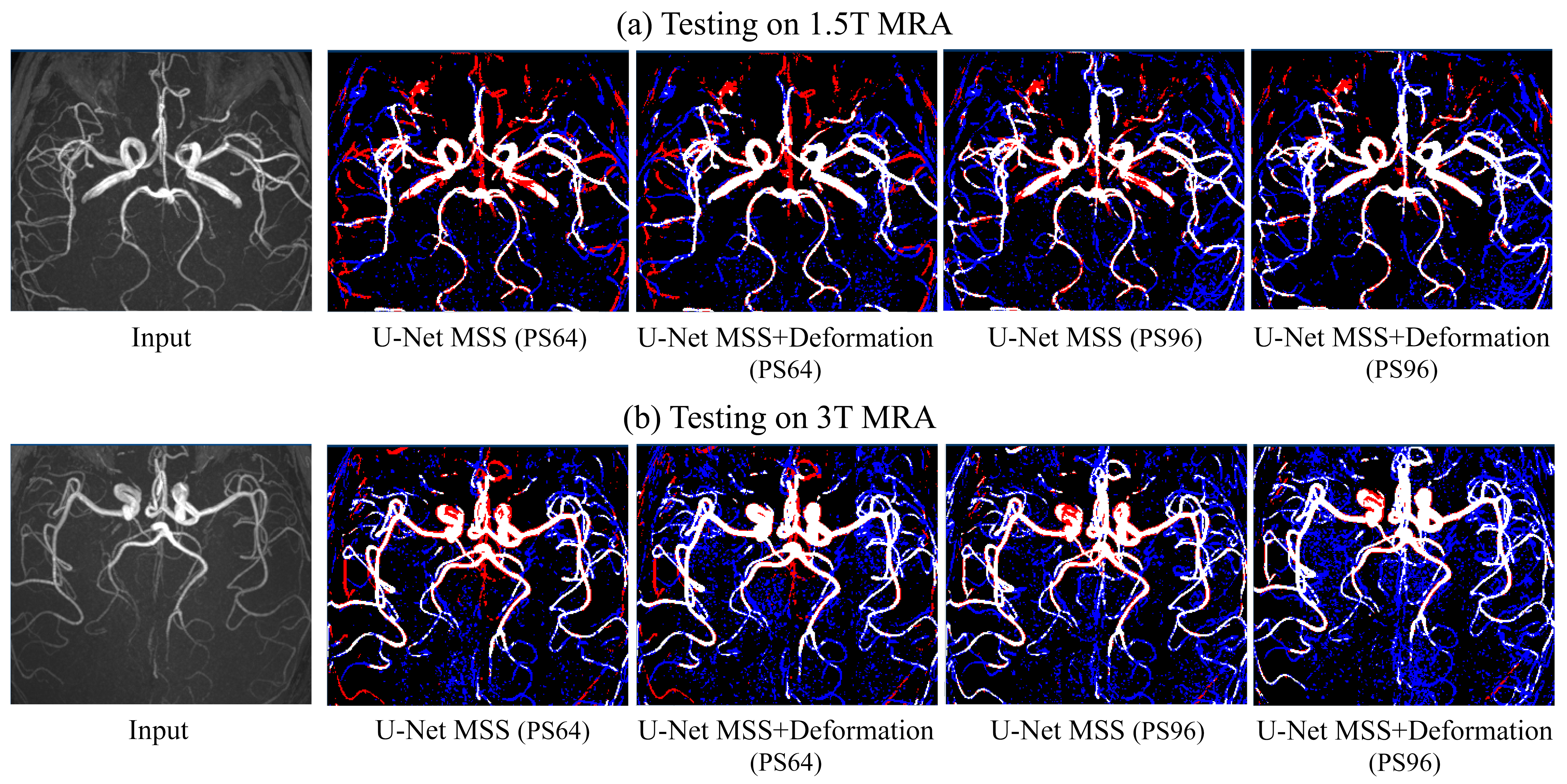}
\caption{Qualitative evaluation of U-Net MSS with and without deformation-aware learning, while segmenting 1.5T and 3T MRA volumes from the IXI dataset. The inference was performed using two different patch sizes: 64 and 96. Red indicates false negative, and blue indicates false positive, while being compared against the dataset labels created with Ilastik.}        
\label{fig8}
\end{figure}

A qualitative analysis was performed on these image sets to gain further understanding of the segmentation performance 
(see Figure~\ref{fig8}). 
 Figure~\ref{fig8}a,b 
 shows the MIP and segmentation mask by U-Net MSS and U-Net MSS + deform with patch sizes 64 and 96 on 1.5T and 3T volumes,
  respectively. The image shows that the predicted segmentation mask is in line with the MIP and shows various areas where the prediction is better than the provided imperfect dataset labels. With the increase in the patch size, there is an improvement in the predictions, but the level of noise increases as well. 
\vspace{-6pt}
\vspace{-10pt}

\begin{table}[H]
\caption{Quantitative evaluation using Dice scores for 1.5T and 3T IXI dataset with different patch~sizes.}\label{tab4}
\newcolumntype{C}{>{\centering\arraybackslash}X}
\begin{tabularx}{\textwidth}{CCCCCC}
\toprule
\multicolumn{2}{c}{} & \multicolumn{2}{c}{\textbf{1.5T}} & \multicolumn{2}{c}{\textbf{3T}}\\
\midrule
\textbf{Model} & \textbf{Patch} & \textbf{Dice Coeff.} & \textbf{IoU} &  \textbf{Dice Coeff.} & \textbf{IoU}\\
\midrule
\multirow{3}{*}{ U-Net MSS}
    & 32 &   { 36.97 \(\pm\) 0.48 }  & {  22.75  \(\pm\)  3.69    } &  { 37.74 \(\pm\) 0.42 }  & {  23.31 \(\pm\)  3,17} \\
    & 64 &   { 39.29 \(\pm\) 0.36 }  & {  24.49  \(\pm\)  2.82    } &  { 40.34 \(\pm\) 0.33}  & {  25.31  \(\pm\)  2.58  } \\
    & 96 &  { 52.51 \(\pm\) 0.82 }  & {  35.88  \(\pm\)  7.44 } &  { 48.25 \(\pm\) 0.16 }  & {  31.81  \(\pm\)  1.47 } \\
\bottomrule
\end{tabularx}
\end{table}

\begin{table}[H]\ContinuedFloat
\caption{{\em Cont.}}
\newcolumntype{C}{>{\centering\arraybackslash}X}
\begin{tabularx}{\textwidth}{CCCCCC}
\toprule
\multicolumn{2}{c}{} & \multicolumn{2}{c}{\textbf{1.5T}} & \multicolumn{2}{c}{\textbf{3T}}\\
\midrule
\textbf{Model} & \textbf{Patch} & \textbf{Dice Coeff.} & \textbf{IoU} &  \textbf{Dice Coeff.} & \textbf{IoU}\\
\midrule
\multirow{3}{*}{U-Net MSS + def}
    & 32 &   { 47.41 \(\pm\) 0.68 }  & {  31.25  \(\pm\)   5.91 } &  { 43.95 \(\pm\) 0.42 }  & {  28.23  \(\pm\) 3.46  } \\
    & 64 &   { 43.94 \(\pm\) 0.39 }  & {  28.22 \(\pm\)   3.31  } &  { 45.96 \(\pm\) 0.23 }  & {  29.86  \(\pm\) 1,97  } \\
    & 96 &  { 58.20 \(\pm\) 0.85 }  & {  41.38 \(\pm\)   8.28   } &  { 49.90 \(\pm\) 0.66 }  & {  33.42\(\pm\) 5.99 } \\
\bottomrule
\end{tabularx}
\end{table}


\section{Discussion}
\label{sec:discussion}

This work presented a self-supervised approach using deformation-aware learning and has
 shown its application to the 
 small vessel segmentation task while being trained with a small, imperfectly labelled dataset. 

 In the above sections, it can be observed that the quantitative, as well as the qualitative performance of U-Net MSS, is higher in comparison to U-Net, as seen in Table~\ref{tab1} and Figure~\ref{fig4}.
  This is in line with the literature (see also the summary of Sections~\ref{section:related_work} and~\ref{section:proposed_methodology})  
  in the sense that the U-Net MSS learns discriminative features as the loss is calculated at multiple levels, and this facilitates efficient gradient flow in comparison to U-Net.  
   Further, it can be observed that the deformation-based models outperform the models trained without deformation (see Table~\ref{tab1}). This could be attributed to the deformation-aware training, where the model is made to learn the shape consistency of the vessels when they are deformed. The findings are in line with the findings of the previously published works~\citep{li2018semi,Bortsova2019}, where it was also shown that making the network transformation equivariant improved the performance.  
   The proposed U-Net MSS model with deformation outperforms all the other models in quantitative as well as qualitative aspects. The authors hypothesise the reason for this to be the efficient gradient flow of U-Net MSS on top of the consistency learning that assists the model in learning discriminative features.

 Given the imperfect training annotation, the authors further strengthen their claim by performing the evaluation against manual segmentation. As seen in Table~\ref{tab3} and Figure~\ref{fig7}, the models trained with deformation-aware learning outperform the baselines. It can be hypothesised that deformation-aware learning helps 
  train the network to be consistent, even when there are inconsistencies present in the annotations.
 Although the manual segmentation is imperfect due to the challenging nature of small vessel segmentation even by a human rater, quantitative estimates and qualitative perception matched well for these ROI-specific comparisons, indicating the improved small vessel segmentation performance of the methods introduced in this study.
 
  While comparing against the noisy dataset labels (created semi-automatically with Ilastik), the modified U-Net MSS performed 4.15\% better than the baseline U-Net. After enabling deformation-awareness for the models, U-Net achieved 4.27\% higher, and U-Net MSS achieved 1.37\% higher Dice than their corresponding models without deformation. As these labels are noisy, the improvements can be observed more clearly when the outputs were compared against manually segmented ground-truth label for an ROI. It is worth stating that the dataset labels created with Ilastik received
   only 50.21 Dice when they were compared against this manually created ROI ground-truth label. While comparing against this manually created label, U-Net MSS achieved 9.95\% better Dice than U-Net. Deformation-aware U-Net got
    26.05\% better than U-Net without deformation, and deformation-aware U-Net MSS got 18.98\% better Dice than the U-Net MSS without deformation. Another important aspect to observe is that when the results were compared to the manual annotation, all the DL methods resulted in less Dice than when they were compared against the semi-automatic Ilastik labels (Tables~\ref{tab1}~and~\ref{tab3},
     respectively)---this can be attributed to the noise in the Ilastik label. However, the non-DL baselines yielded better Dice scores while comparing against manual annotations than while comparing against the Ilastik labels. Moreover, in these comparisons against the manual labels, the non-DL baselines yielded better Dice than the DL models without deformation. This can be attributed to the fact that the models which were trained using imperfect labels and without deformation-awareness failed to learn properly.   
 
 The performance of the model trained with and without deformation was analysed on lower resolution image volumes (450 \textmu m isotropic resolution) of 1.5T and 3T, where the model results improved with increasing patch size and was able to capture vessels that were not captured in the annotation but the level of noise in the results increased as well. The network was trained using image volumes with 300 \textmu m isotropic resolution with a patch size of 64, and it performed best for patch size 96 for the 450 \textmu m image volumes. 
 So,
  it can be said that the performance of the model was dependent on the patch size, and a clear relationship can be observed between the image resolution and patch size, as when the image resolution decreased by a factor of 1.5, the patch size had to be increased by a factor of 1.5 to get better results. 
  This is counter-intuitive and demands further in-depth investigations.  
   It is to be noted that the models were trained on 7T MRA without bias field correction, and due to that fact, the
    7T volumes had field inhomogeneity. 
     The 1.5T and 3T~volumes did not have such observable inhomogeneity, and testing models trained on 7T~data might have had a negative impact due to this fact.
 
 The experiments performed with the different training set sizes have shown that the trainings with six volumes performed better than the trainings performed with smaller datasets. However, trainings performed with two and four volumes were inconclusive. They resulted in similar Dice scores for the proposed method (77.84 $\pm$ 2.35 and 77.79~$\pm$~2.05~Dice, respectively, 
 for U-Net MSS + Deformation), but a training set size of two performed better than four for the rest of the models, which is completely counter-intuitive and further investigations are needed to determine the cause. Trainings performed with one volume yielded poorer, but still acceptable results (74.81 $\pm$ 1.32 Dice for the proposed U-Net MSS  + Deformation). One possible future experiment can be performed by creating one perfect ground-truth for one volume and training the models with just that volume, and comparing the performance against this paper where up to six volumes (imperfectly labelled) were used for training. 
 
 The limitation of the proposed approach is the missing continuity of the small vessels in some of the regions, as seen in the last column of Figure \ref{fig4}. This can be attributed to the fact that the model was trained using patches and lacks the context of continuity within the whole volume, which leads the model to predict stray points with a higher intensity as small vessels, which could be avoided if continuity is taken into consideration. Prior knowledge of vessels being cylindrical in nature is disrupted with deformation to a certain extent. 
 In addition,
  since the Dice coefficient is calculated concerning an imperfect labelled image, some of the inherently detected vessels in the proposed model are considered over-segmented.  
   
 The main challenge for the quantitative evaluation is the missing ground truth due to the non-trivial nature of (small) vessel segmentation itself as well as the study design, i.e., learning from imperfect annotations.
 Therefore, the quantitative metrics reported here (i.e., Dice and IoU) have to be interpreted with care. Hence, a qualitative comparison of whole-brain and ROI-specific MIPs were used to strengthen any claims made from the quantitative assessment. ROI-specific MIPs were further used to judge small vessel segmentation qualitatively. Further, for a single ROI, manual segmentation was done to further strengthen claims on small vessel segmentation quantitatively and qualitatively (see Figure~\ref{fig7}). 
 Imperfection in the annotations as well as manual segmentation of a single ROI prevented a detailed quantitative comparison of small vs medium to large-scale segmentation performance by,
  i.e.,
   stratifying the vessels by their diameter. This is mainly due to false positives, i.e., noise in the segmentation and vessel discontinuities which would be a considerable bias in any diameter-specific comparison of segmentation methods.

 Furthermore, the models were evaluated for robustness against different MR image artefacts by performing artefact testing---shown in Appendix~\ref{app:artefacts}. 
 It was revealed that the models are not robust to random spikes but are robust against random noise and random bias fields.
 

\section{Conclusions}
The proposed network uses a modified version of U-Net Multi-Scale Supervision (MSS) architecture with deformations, which makes it equivariant to elastic deformations. The evaluation of this proposed approach was performed for the task of small vessel segmentation from MR angiograms. The training and testing of the model were performed using a 7T MRA dataset with a resolution of 300 \textmu m. Segmentation of 7T MRA is highly relevant as it can help in the detection of cerebral small vessel disease. The proposed method was compared against non-machine learning methods, such as the Frangi filter and the MSFDF pipeline, as well as against Deep learning methods, such as U-Net and Attention U-Net. The proposed method outperformed all these baseline methods while being trained using a small, imperfectly labelled (noisy) dataset. It could be observed that the models with deformation consistency significantly outperformed the ones without. When the models were trained with only one volume, deformation-aware models could maintain vessel continuity better than their counterparts without deformation-aware learning. As the dataset labels were imperfect, the improvements achieved by the proposed model (U-Net MSS + Deformation) can be observed clearer when the results were compared against manually segmented ROI, where U-Net MSS achieved 52.17 and 62.07 Dice scores, with and without deformation, respectively. This research has shown that the proposed modified U-Net MSS with deformation-aware learning outperformed the baseline methods, and deformation-aware learning outperformed models without deformation-awareness. Furthermore, it can be said that the current trained network of the proposed method can be employed to perform vessel segmentation on 7T MRA. The network can be fine-tuned and also used for 1.5T and 3T MRA images. The possibility of training the proposed DS6 approach on a 
small dataset for the task of vessel segmentation has been demonstrated here. This might also be applicable for other tasks to combat the small training dataset problem. 
 The code of this research is publicly available on GitHub 
 \url{https://github.com/soumickmj/DS6} - Accessed on 21.09.2022). 

In future work, the authors aim to explore further generalisability of the model by performing mixed training with different resolutions (150 \textmu m, 300 \textmu m and 450 \textmu m) and with different field strengths (1.5T, 3T and 7T). Moreover, trainings will be performed using MIP as an additional loss term to try to improve the continuity of vessels. Experiments can also be performed by training the network from scratch by using the network's predicted output as labels, as they are better than the original imperfect annotations. Furthermore, introspection on how deformation-aware learning is helping the models learn different shapes of vessels is also another interesting future direction of work. Finally, the superiority of the proposed method while training on a small dataset and its applicability in segmentation tasks with different sizes of regions of interest, as observed in this research, makes it interesting to apply to
 other image segmentation tasks where requirements are similar.


\vspace{6pt} 



 \authorcontributions{Conceptualisation, S.C., H.M., F.D., and O.S.; methodology, S.C., K.P., M.P., G.B., F.D., and A.N.; software, S.C., K.P., and M.P.; validation, S.C., C.S., and H.M.; formal analysis, S.C., H.M., and F.D.; investigation, K.P. and M.P.; resources, H.M., O.S., and A.N.; data curation, H.M., K.P., and M.P.; writing---original draft preparation, S.C.; writing---review and editing, G.B., F.D., H.M., M.d.B., O.S., and A.N.; visualisation, C.S.; supervision, M.d.B., O.S., and A.N.; project administration, A.N.; funding acquisition, M.d.B., O.S., and A.N. All authors have read and agreed to the published version of the manuscript.}

\funding{Soumick Chatterjee and Chompunuch Sarasaen were funded within the context of the International Graduate School MEMoRIAL at Otto von Guericke University (OVGU) Magdeburg,
Germany, kindly supported by the European Structural and Investment Funds
(ESF) under the program Sachsen-Anhalt WISSENSCHAFT Internationalisierung (project no. ZS/2016/08/80646).
Gerda Bortsova was funded by the Dutch Technology Foundation STW, which is part of the Netherlands Organisation for Scientific Research (NWO), project number P15-26, and co-funded by Intel Corporation. 
Florian Dubost was funded by The Netherlands Organisation for Health Research and Development (ZonMw) Project 104003005. Hendrik Mattern was supported by the DFG 
(grant number MA 9235/1-1).}

\institutionalreview{Ethical review and approval were waived for this study because only retrospective analyses were performed. The main dataset used in this study has been published by Mattern et al. (in 2018, Prospective motion correction enables highest resolution time-of-flight angiography at 7T. Magnetic Resonance in Medicine 80, 248–258) and the additional dataset is publicly available. }

\informedconsent{Informed consent was obtained prior to being scanned from all the subjects involved in the study for the acquisition of the original manuscript of the data (Mattern et al. 2018. Prospective motion correction enables highest resolution time-of-flight angiography at 7T. Magnetic Resonance in Medicine 80, 248–258).}

\dataavailability{The training data, as well as the data used for the main experiments of this paper, are not publicly available due to data privacy and security concerns regarding medical data. The manuscript of the dataset is: “Mattern et al. 2018. Prospective motion correction enables highest resolution time-of-flight angiography at 7T. Magnetic Resonance in Medicine 80, 248–258.  \url{https://doi.org/10.1002/mrm.27033}”. Data might be available on request from the corresponding author of that original manuscript. The annotations may also be made available as part of the same request. Additional evaluations were performed using the MRA volumes from the IXI dataset, which is publicly available at the link: \url{https://brain-development.org/ixi-dataset/} (Accessed on 21.09.2022)} 

\acknowledgments{The authors would like to acknowledge the contribution of Harish Kumar Harihara Subramanian of the Faculty of Computer Science, Otto von Guericke University Magdeburg, Germany, for his efforts with data handling and label creation. }

\conflictsofinterest{The authors declare no conflicts
 of interest.} 





\appendixtitles{yes} 
\appendixstart
\appendix
\section[\appendixname~\thesection. Testing with Artefacts]{Testing with Artefacts}
\label{app:artefacts}
Different types of artefacts may be present in acquired MR images. Testing the performance of the models in the presence of these artefacts can help understand the robustness of the model against such artefacts. The authors used the standard artefact transformations provided by the TorchIO~\citep{fern2020torchio} library. The different artefacts listed in Table~\ref{tabA1} were applied to the volumes before supplying them to the models during inference.

\begin{figure}[H] 
\includegraphics[width=0.85\textwidth]{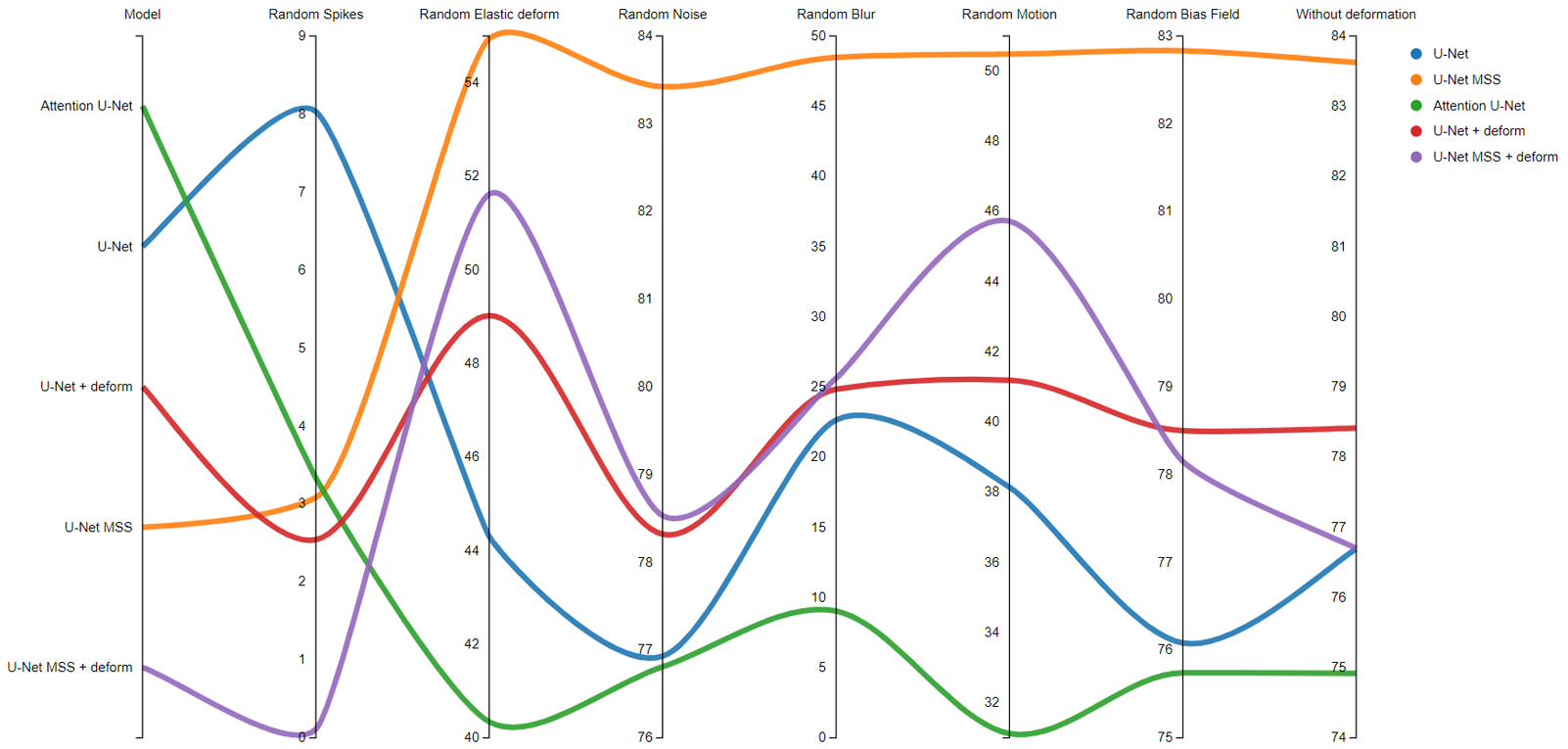}
\caption{Resultant average Dice scores yielded by the different models after receiving input volumes having different types of MR artefacts.}

\label{figA1}
\end{figure}

Figure~\ref{figA1} showcases the response of the models to different artefacts. It is evident that for random spikes and random motion, all the models have reported unsatisfactory performance, which means none of these models is robust against these artefacts. Compared to the other artefacts, elastic deformation and random noise result only in slightly reduced segmentation performance. Lastly, an interesting behaviour of the models can be observed against random blur because the Attention U-Net model displays exceptionally bad performance in comparison to other models. 

\begin{figure}[H]
\begin{tabular}{ccc}
    \includegraphics[angle=180,origin=c,width=.23\linewidth]{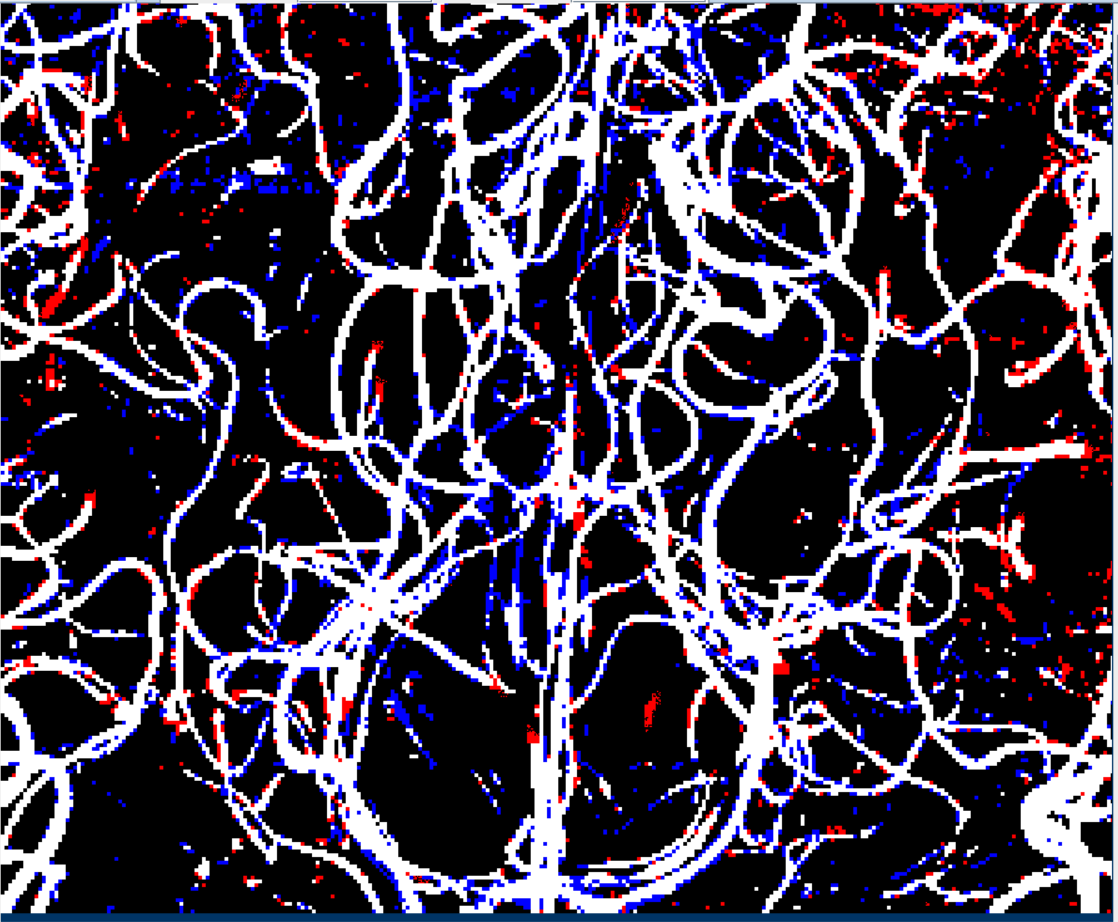} &
    \includegraphics[angle=180,origin=c,width=.23\linewidth]{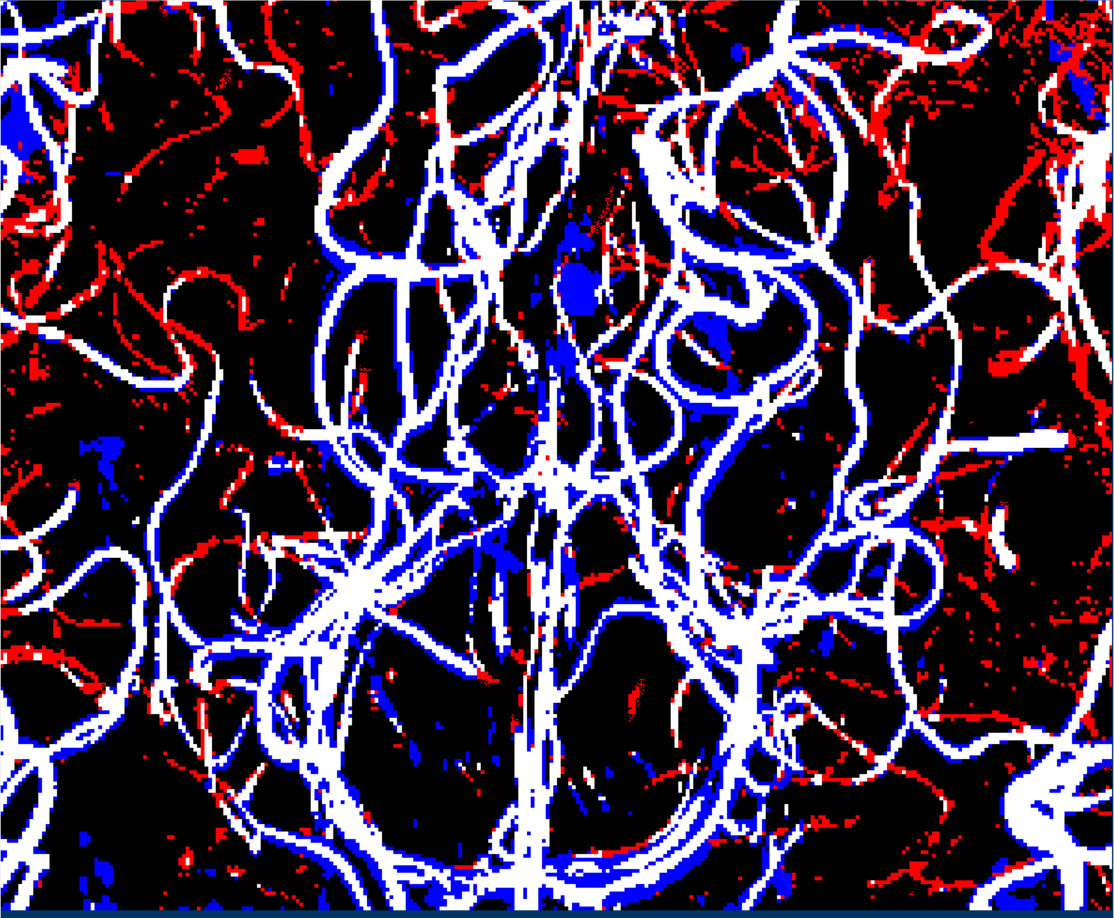} &
    \includegraphics[angle=180,origin=c,width=.23\linewidth]{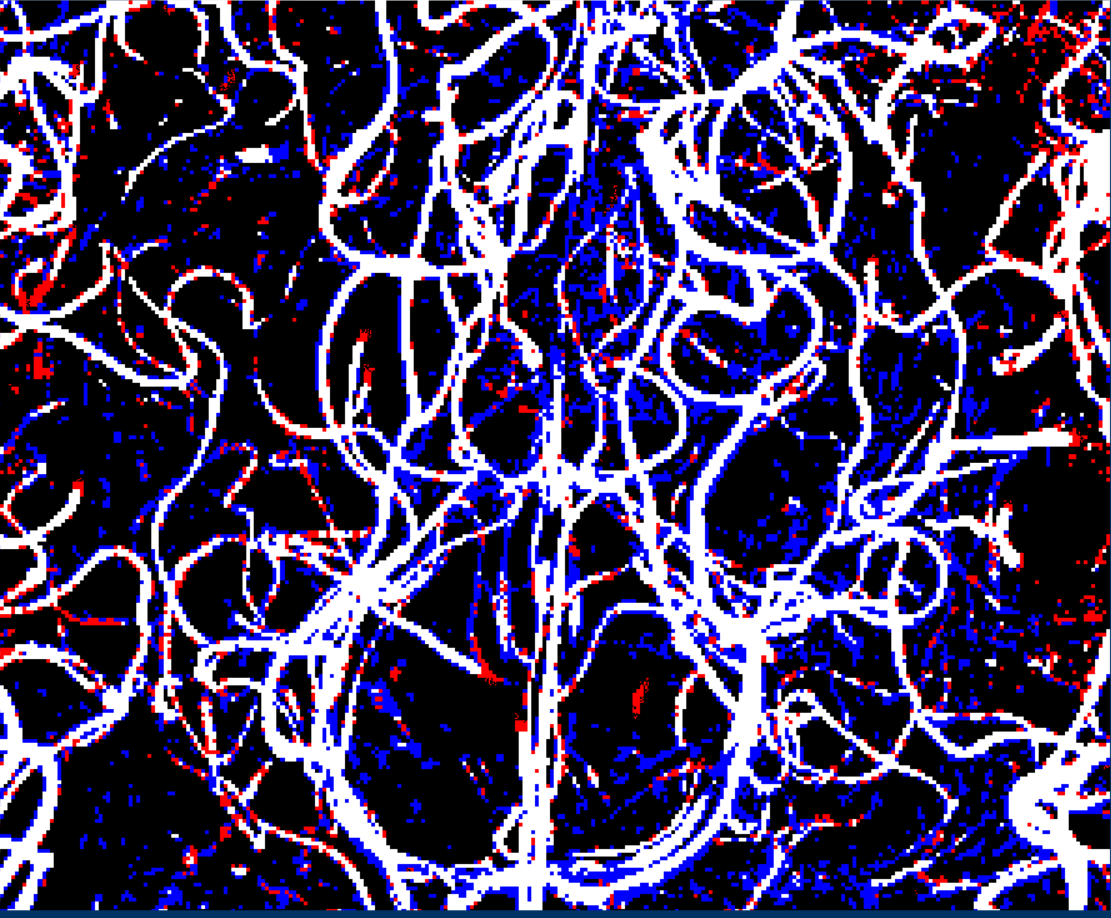} \\
    \includegraphics[angle=180,origin=c,width=.23\linewidth]{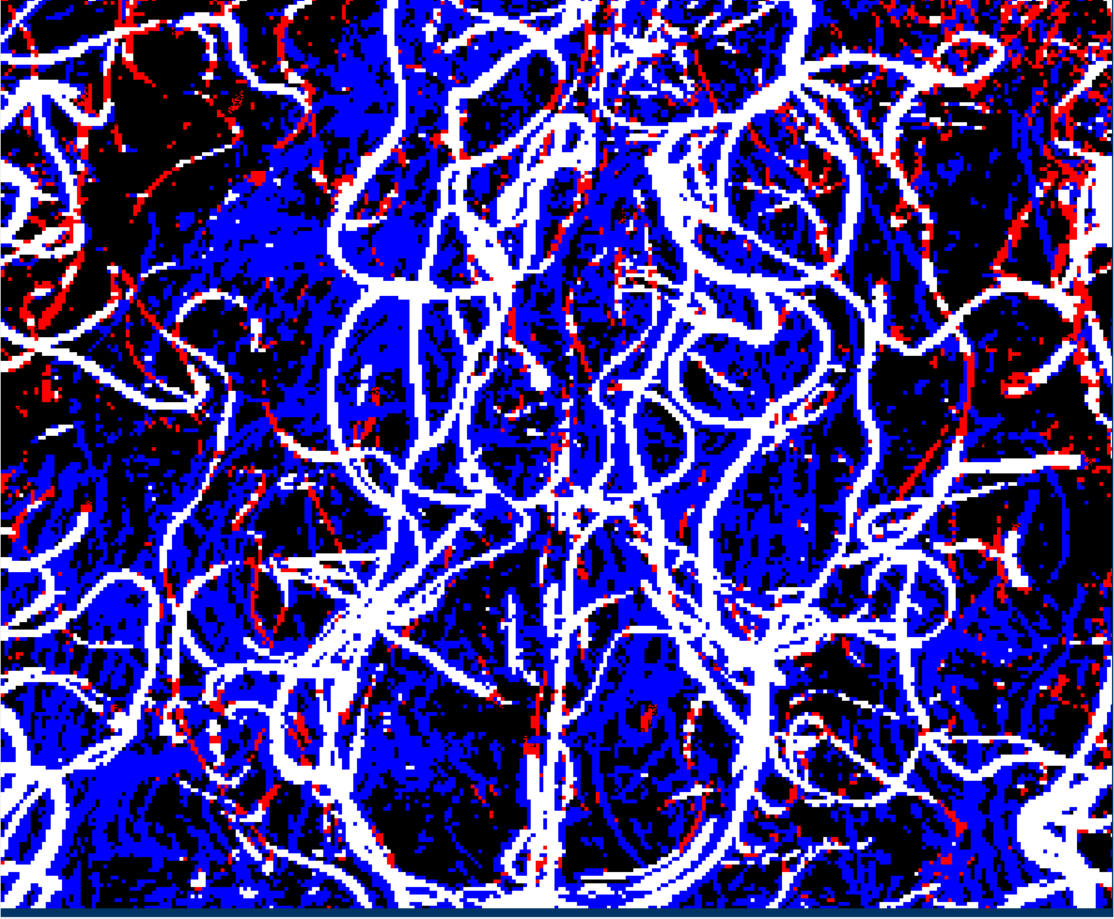} &
    \includegraphics[angle=180,origin=c,width=.23\linewidth]{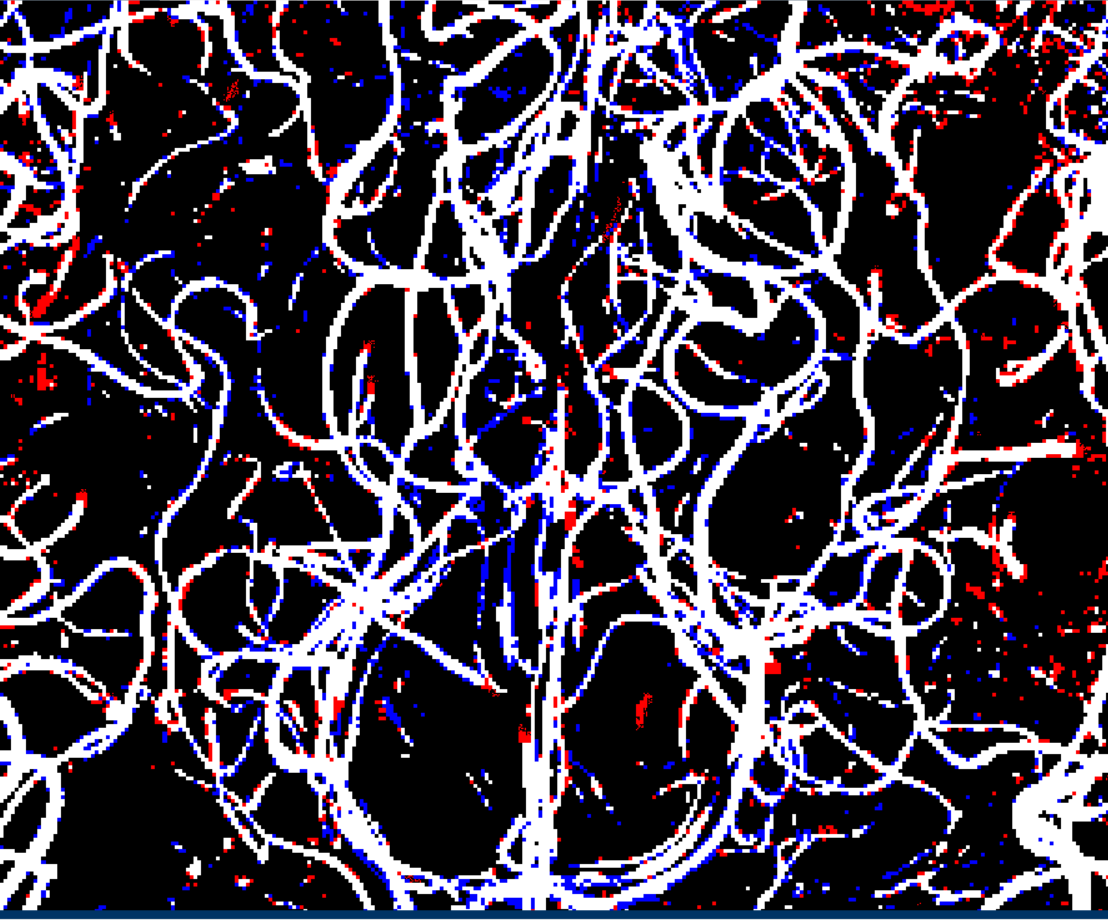}&
    \includegraphics[angle=180,origin=c,width=.23\linewidth]{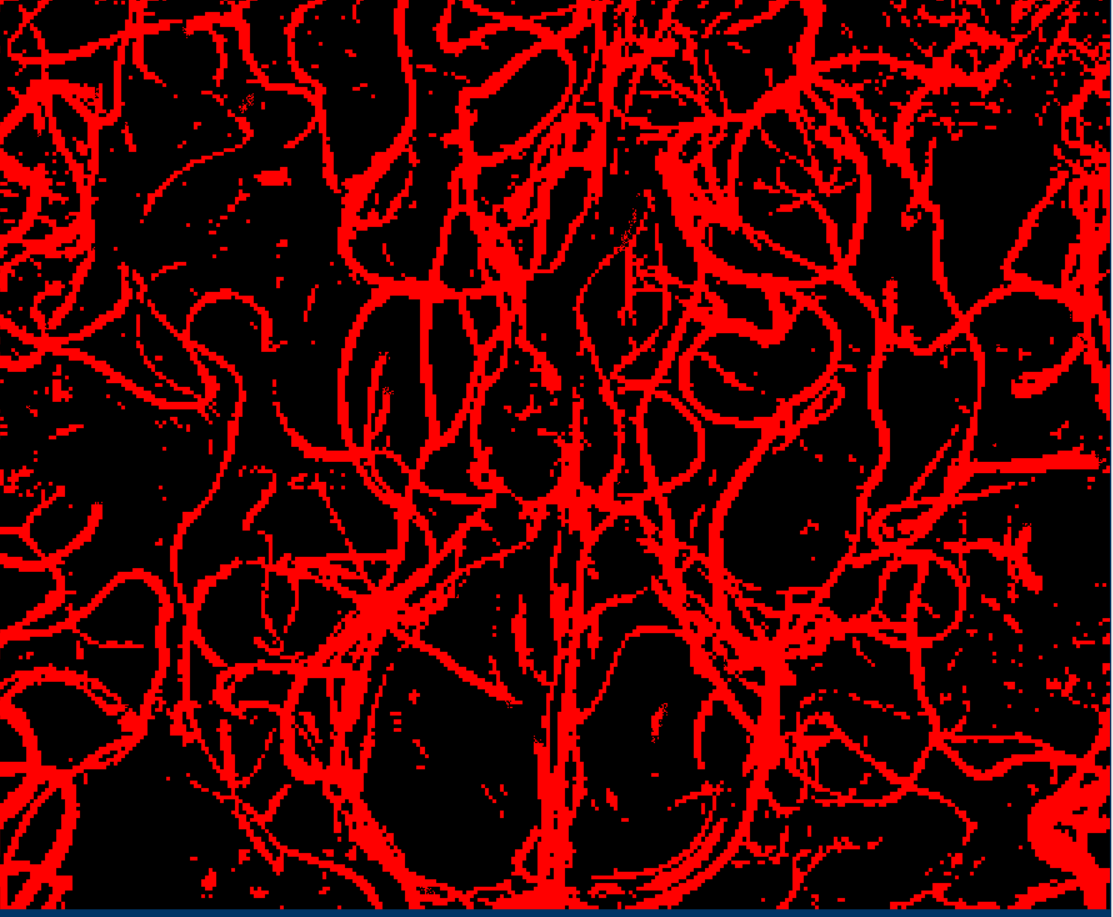}\\
    \multicolumn{3}{c}{({\bf a})}\\
  
    \includegraphics[angle=180,origin=c,width=.23\linewidth]{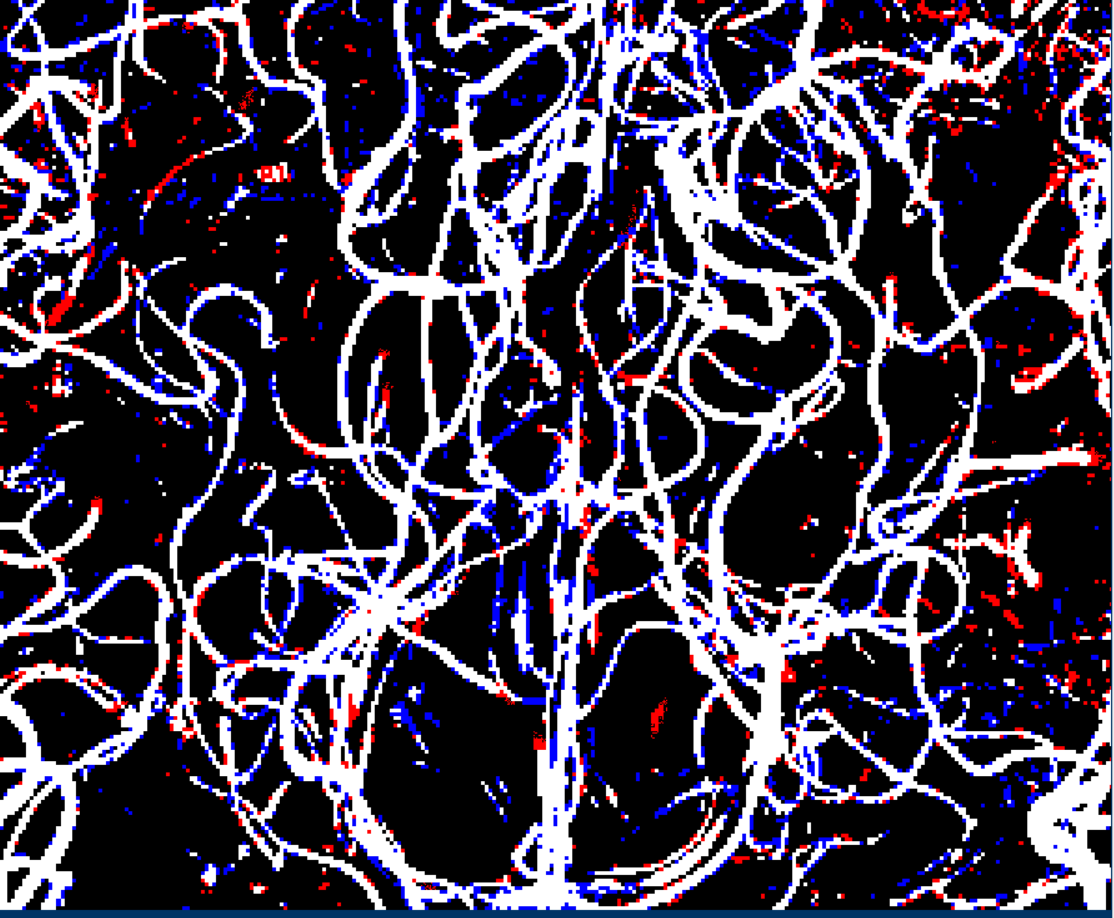} & 
    \includegraphics[angle=180,origin=c,width=.23\linewidth]{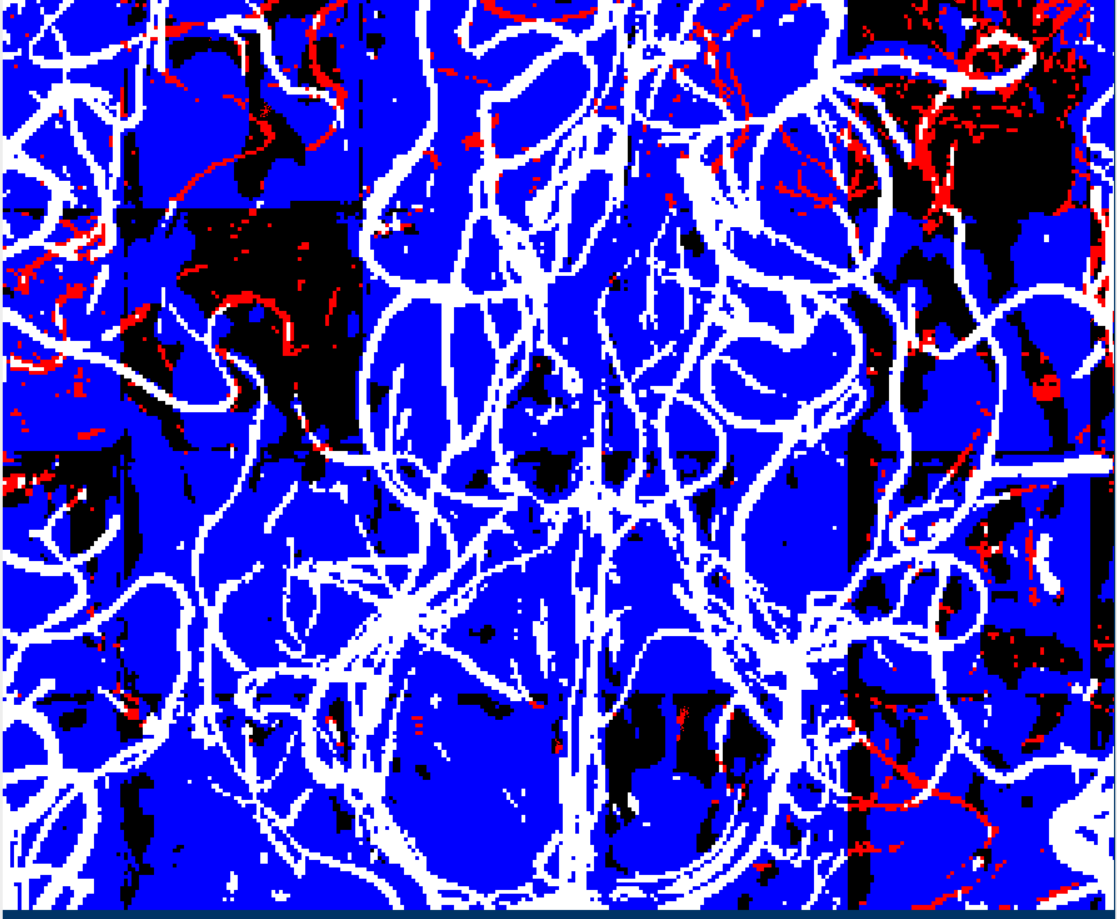}  &
    \includegraphics[angle=180,origin=c,width=.23\linewidth]{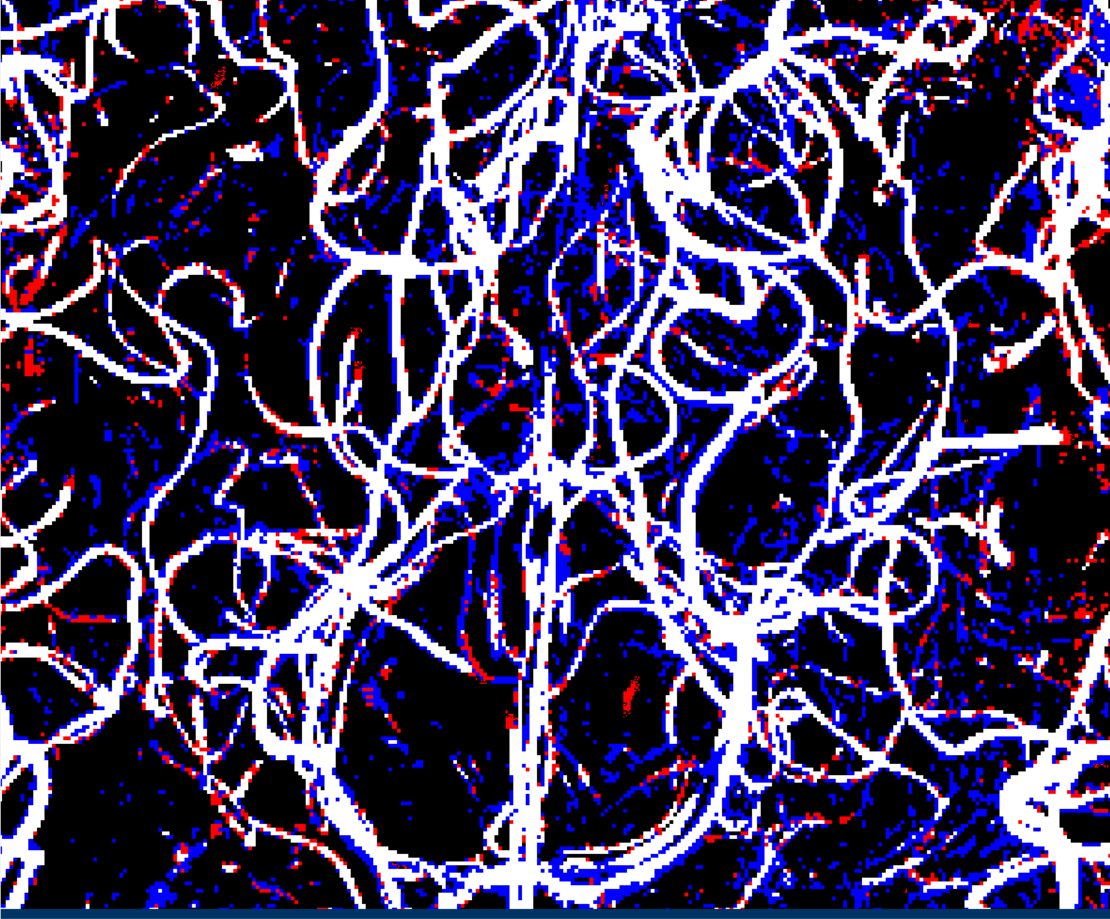} \\
    
    \includegraphics[angle=180,origin=c,width=.23\linewidth]{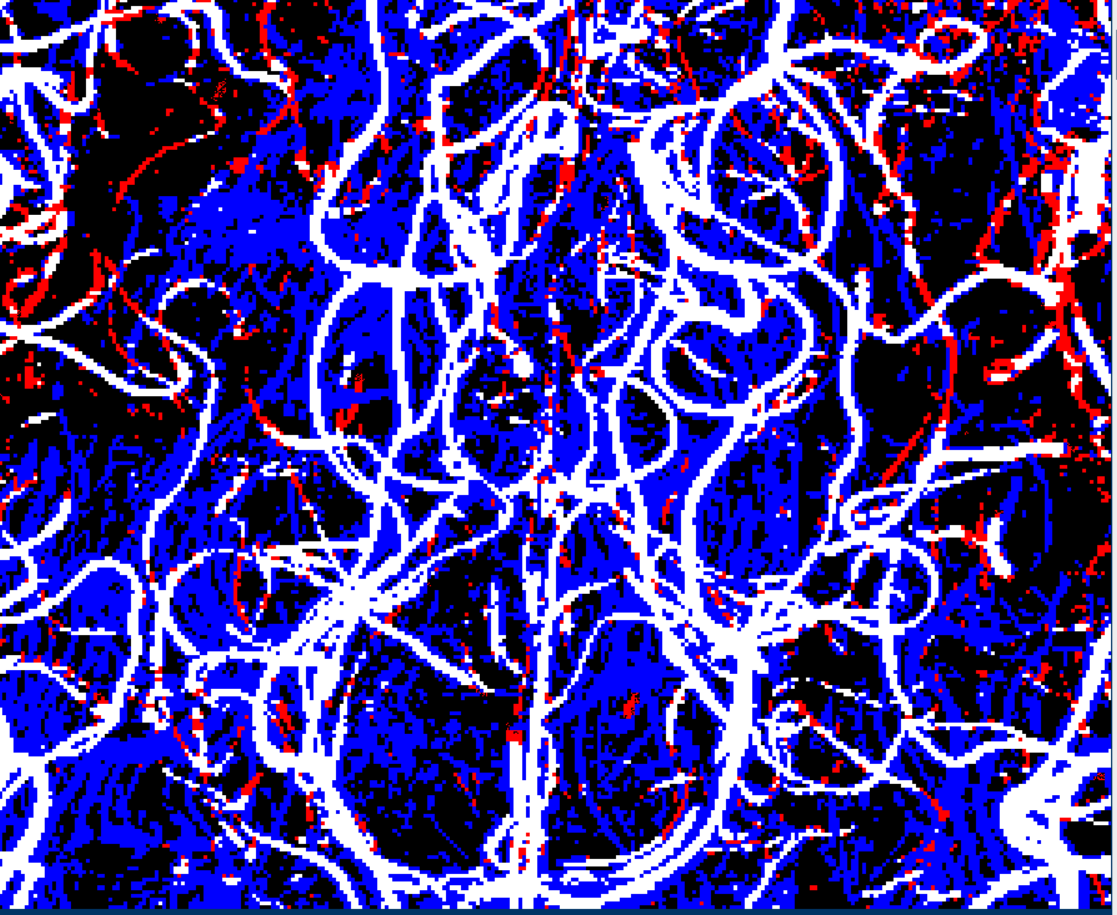} &
    \includegraphics[angle=180,origin=c,width=.23\linewidth]{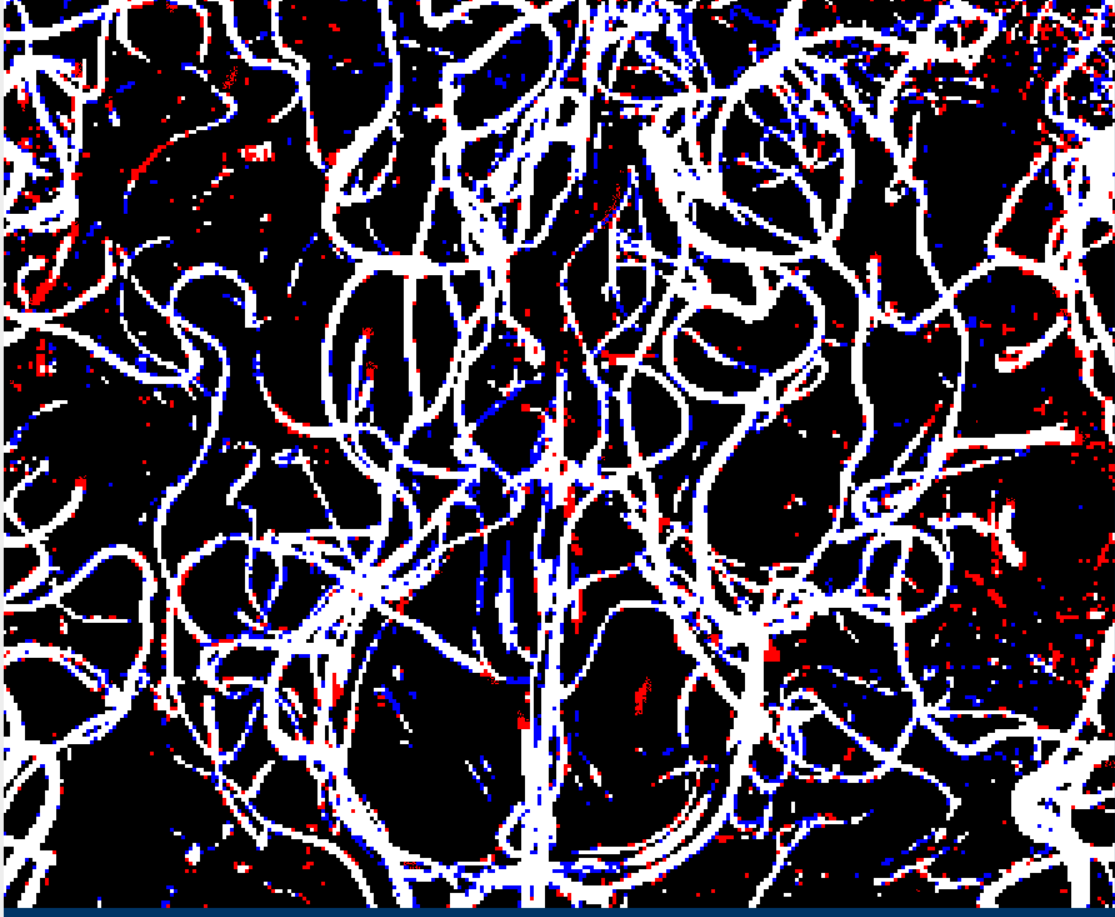} &
   \includegraphics[angle=180,origin=c,width=.23\linewidth]{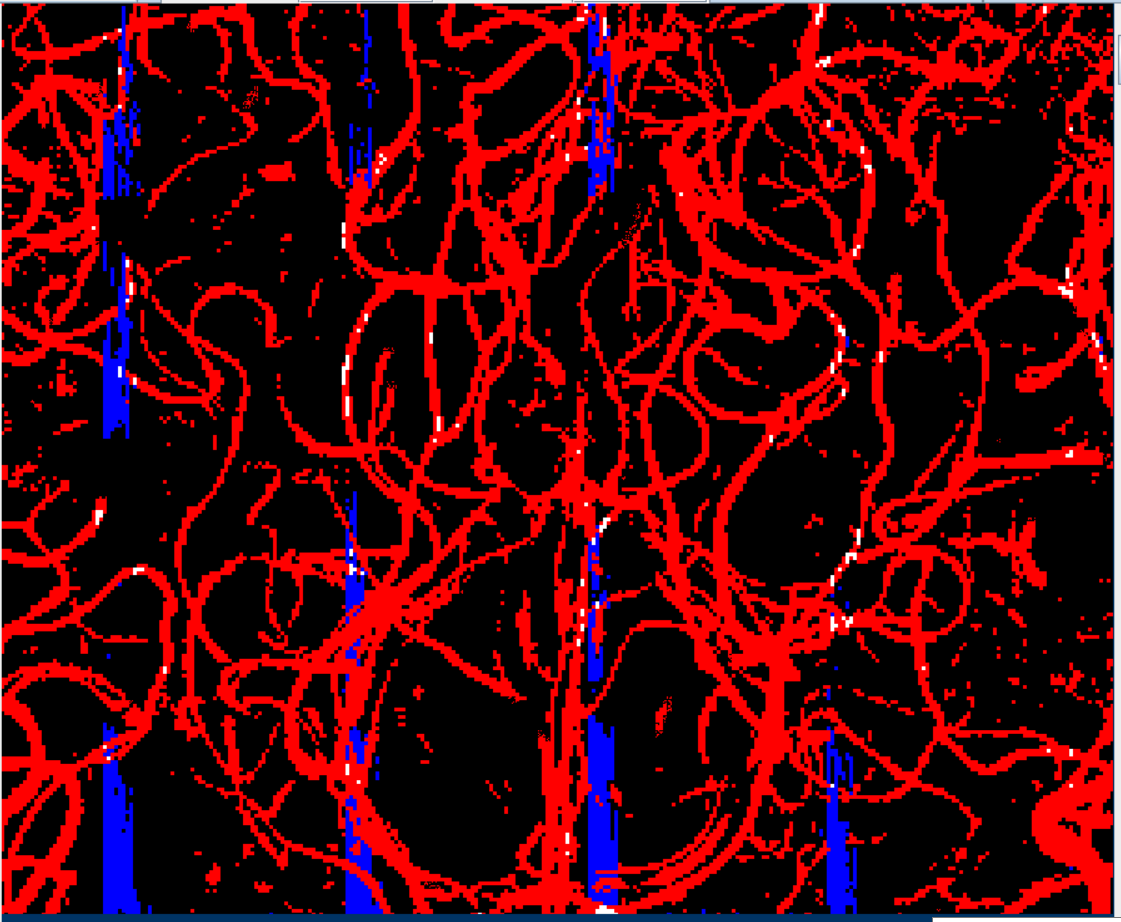} \\
    \multicolumn{3}{c}{({\bf b})}\\

    \includegraphics[angle=180,origin=c,width=.23\linewidth]{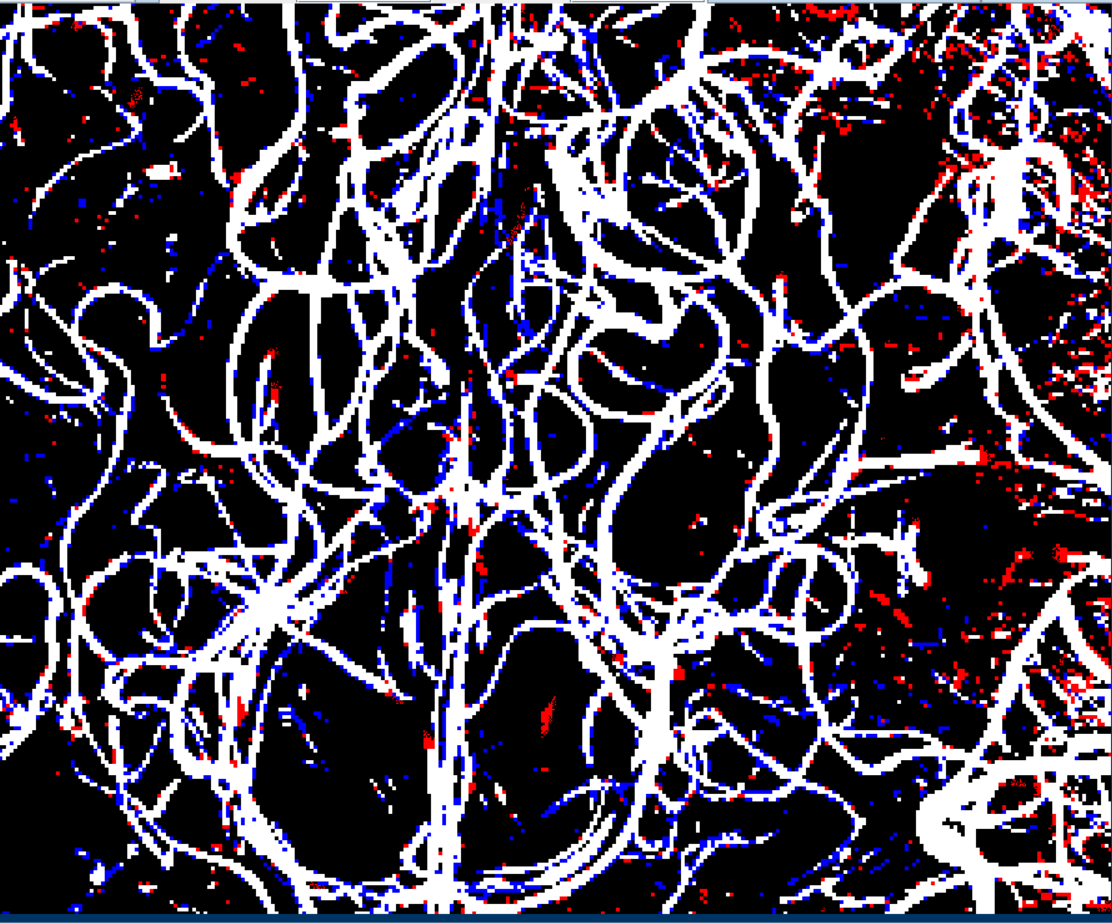} &
    \includegraphics[angle=180,origin=c,width=.23\linewidth]{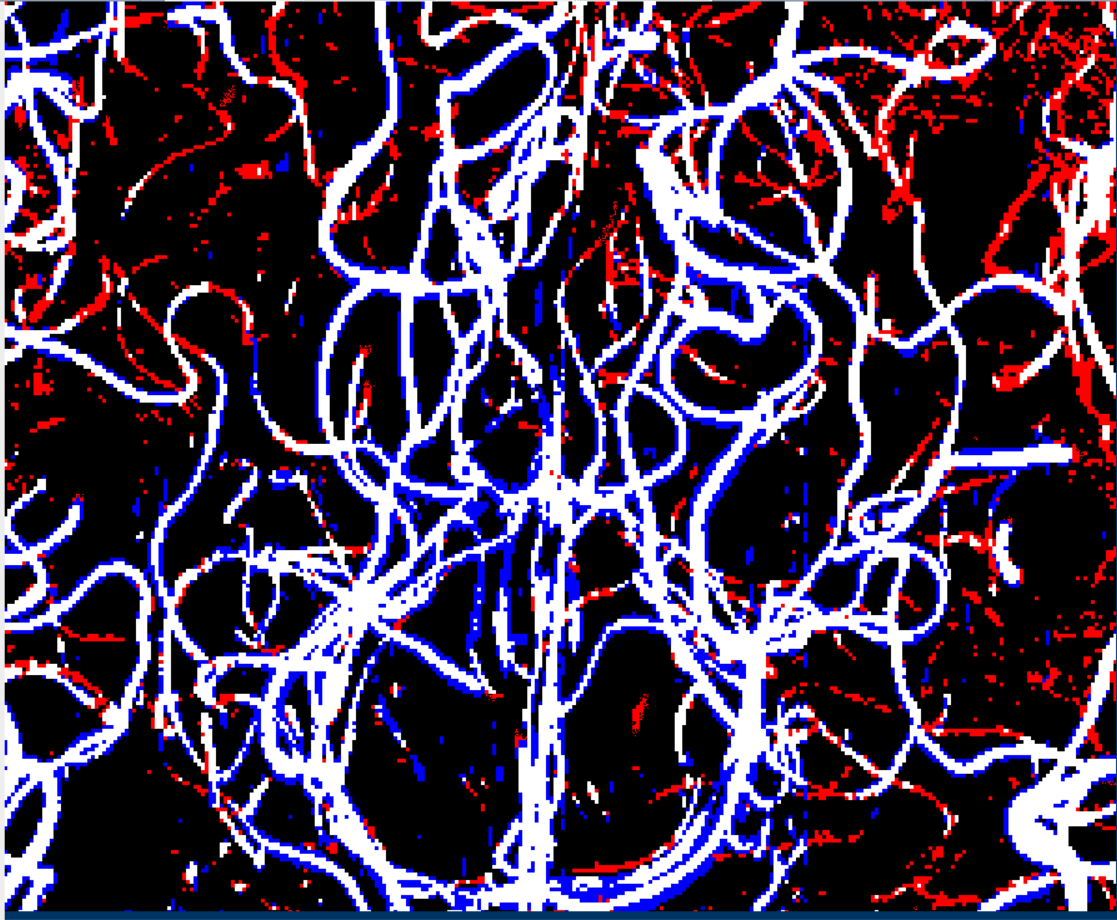} &
    \includegraphics[angle=180,origin=c,width=.23\linewidth]{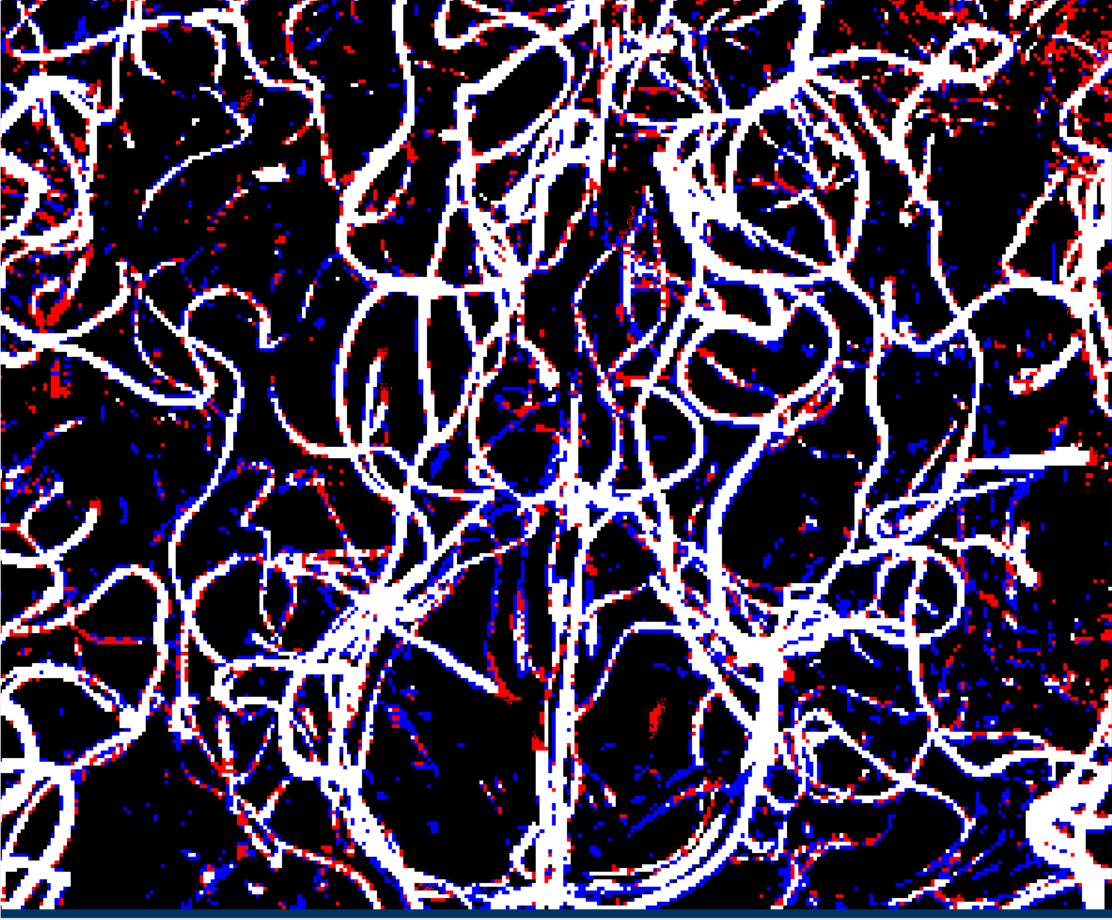} \\
  
    \includegraphics[angle=180,origin=c,width=.23\linewidth]{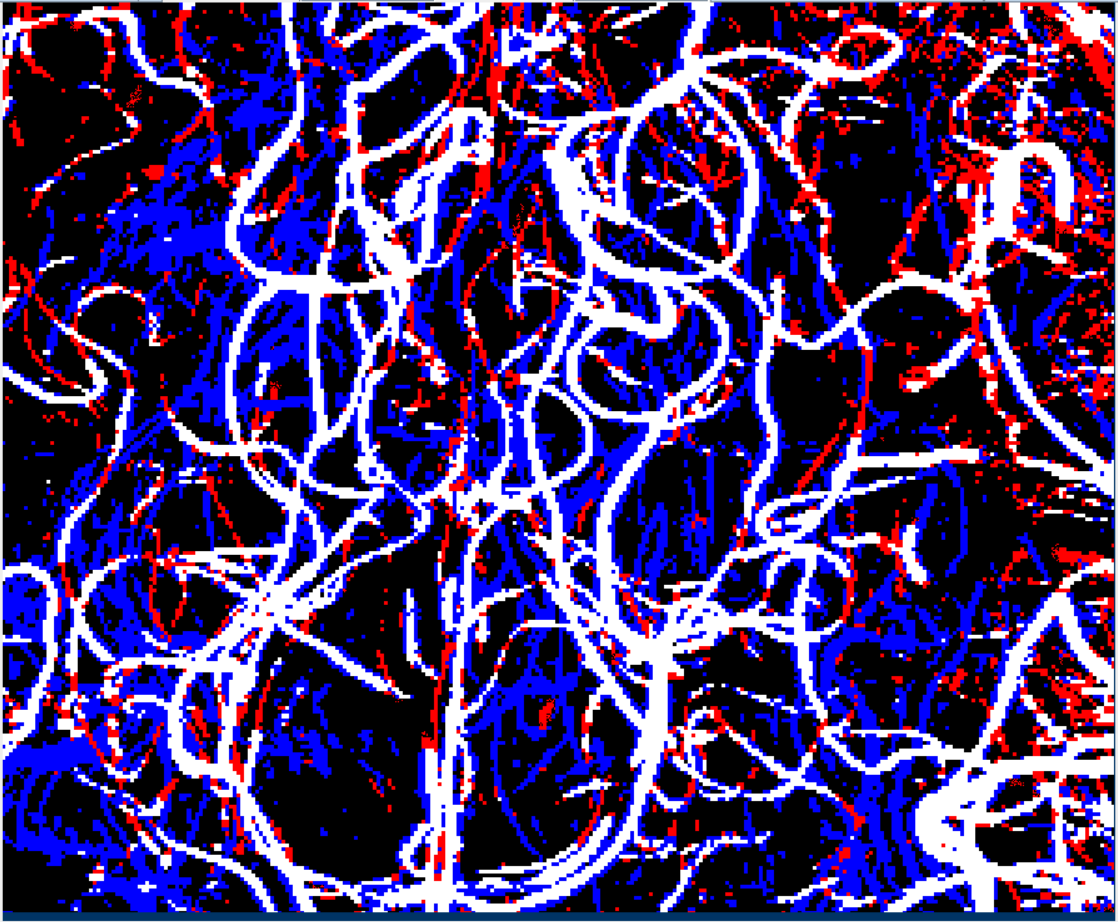} &
    \includegraphics[angle=180,origin=c,width=.23\linewidth]{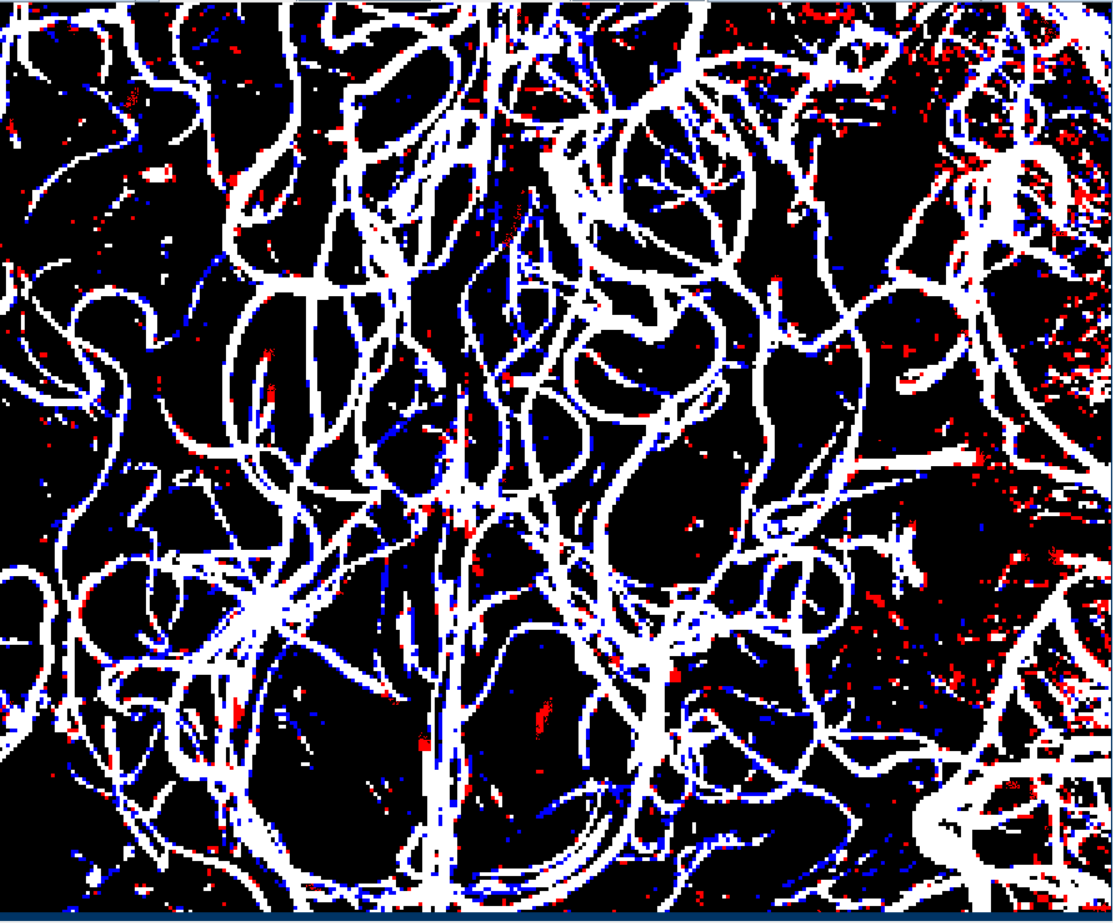} &
    \includegraphics[angle=180,origin=c,width=.23\linewidth]{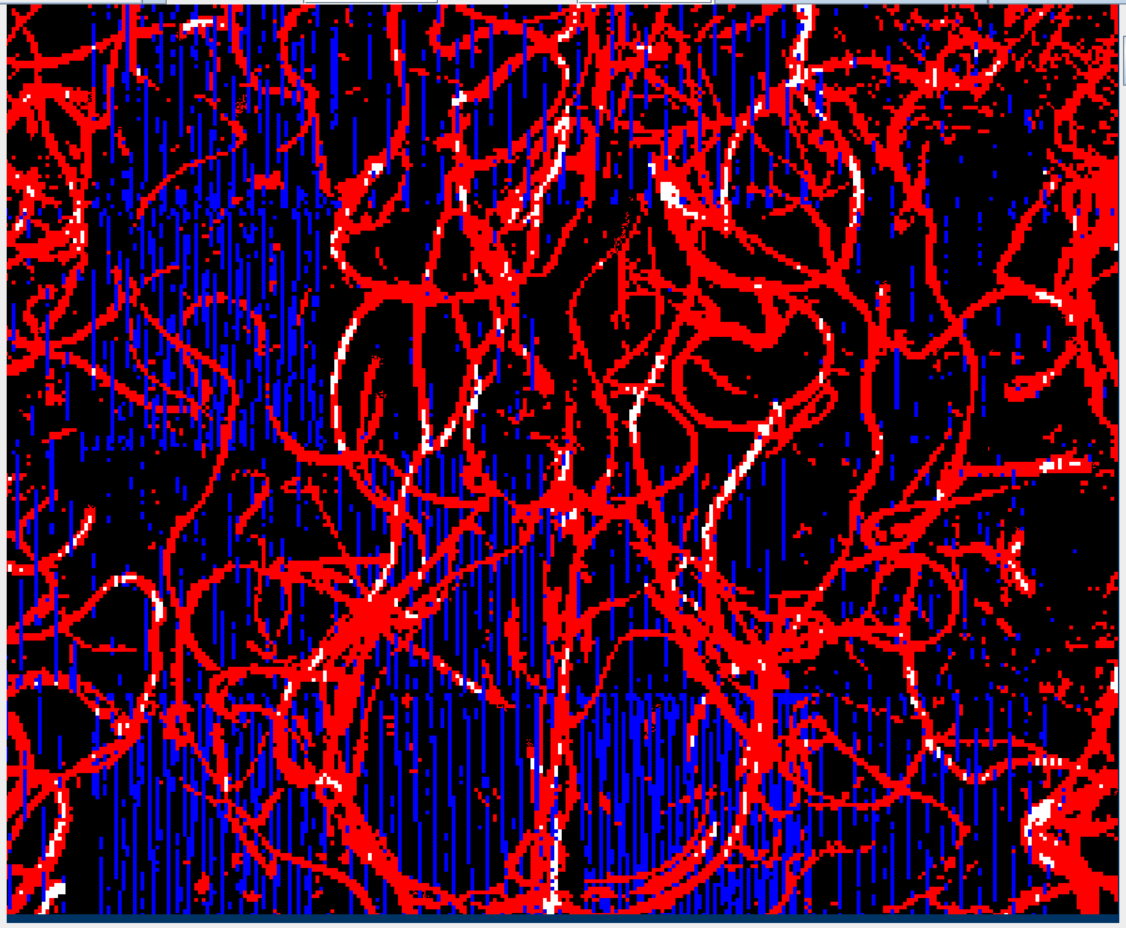} \\
    \multicolumn{3}{c}{({\bf c})}\\
    \end{tabular}

\caption{Performance 
 of the models \textit{without 
} deformation-aware learning, while dealing with various MR artefacts: (First row left to right) Random Bias Field, Random Blur, Elastic deformation (Second row left to right) Random Motion, Random Noise, and Random Spike. Red indicates false negative and blue indicates false positive while comparing against the dataset labels. (\textbf{a}) U-Net. (\textbf{b}) Attention U-Net. (\textbf{c}) U-Net MSS.}        
\label{figA2}
\end{figure}

\begin{figure}[H]
\begin{tabular}{ccc}
    \includegraphics[angle=180,origin=c,width=.23\linewidth]{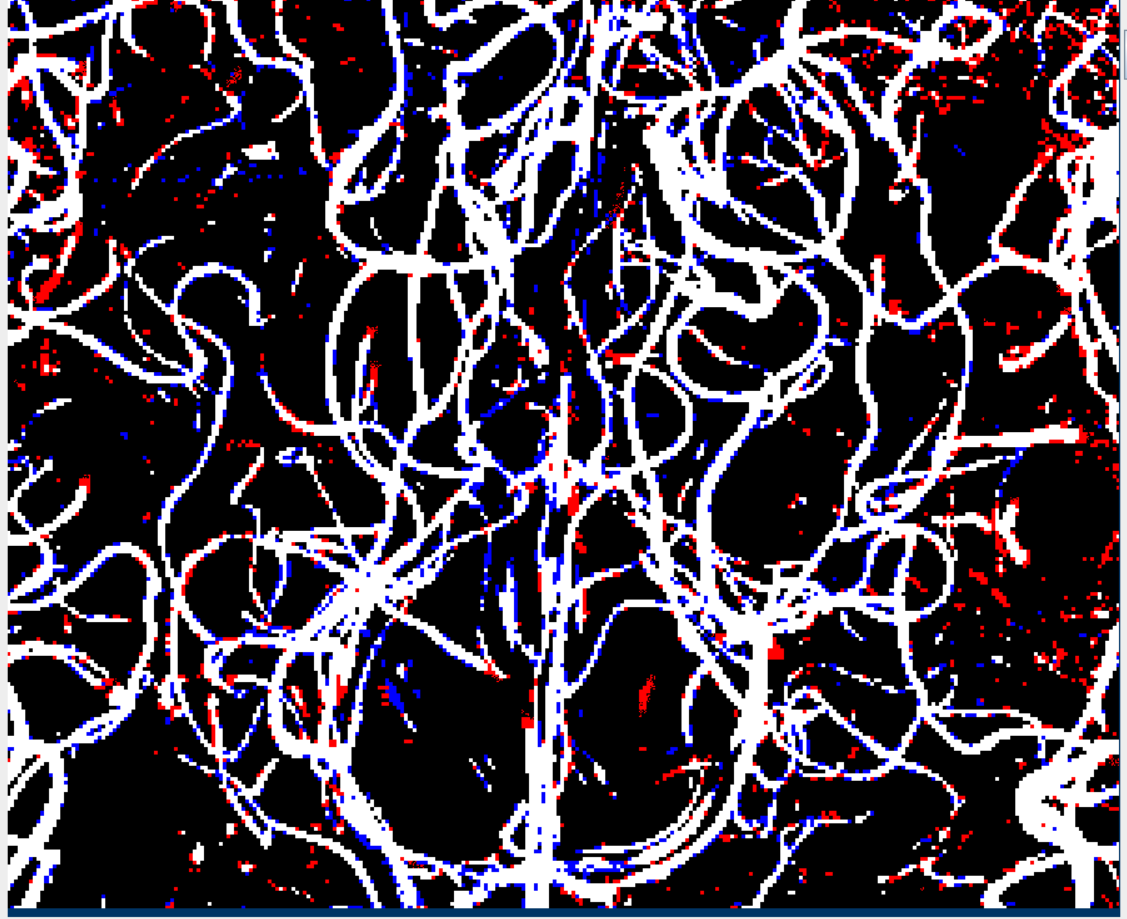} &
    \includegraphics[angle=180,origin=c,width=.23\linewidth]{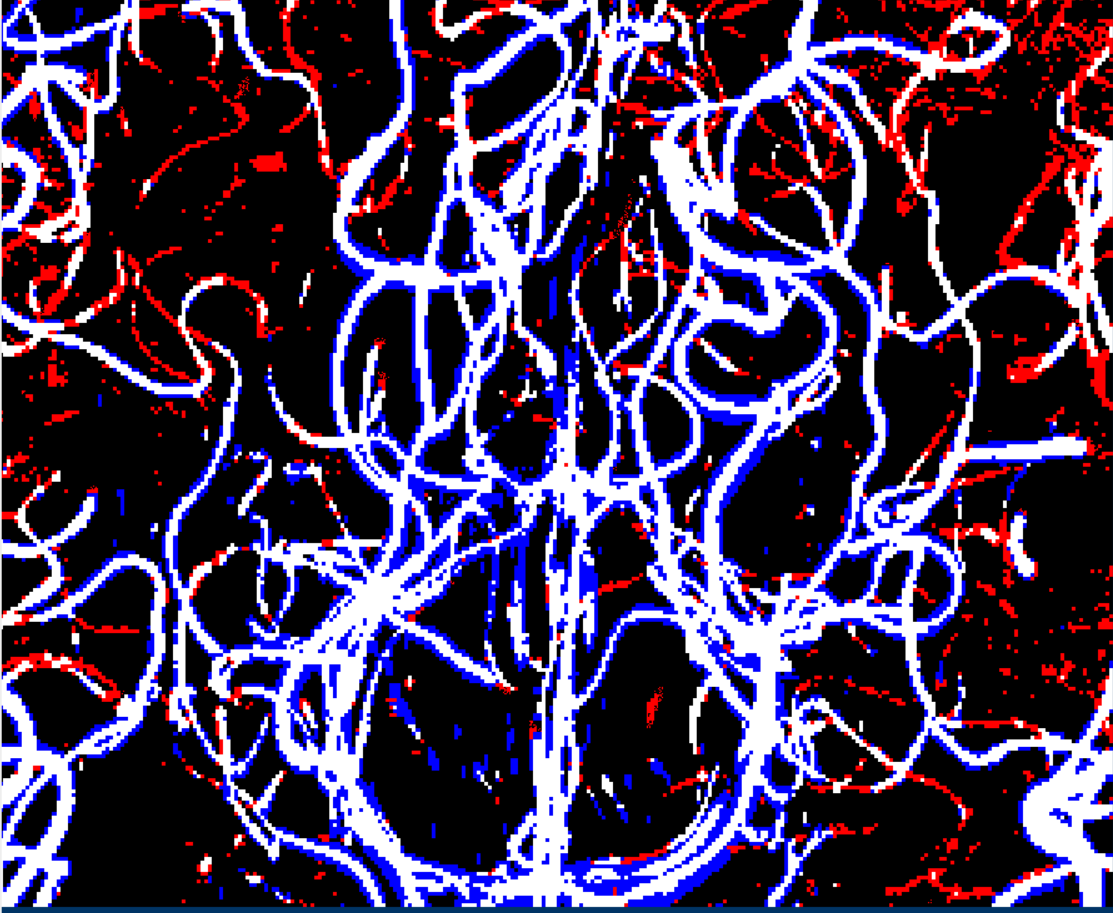} &
    \includegraphics[angle=180,origin=c,width=.23\linewidth]{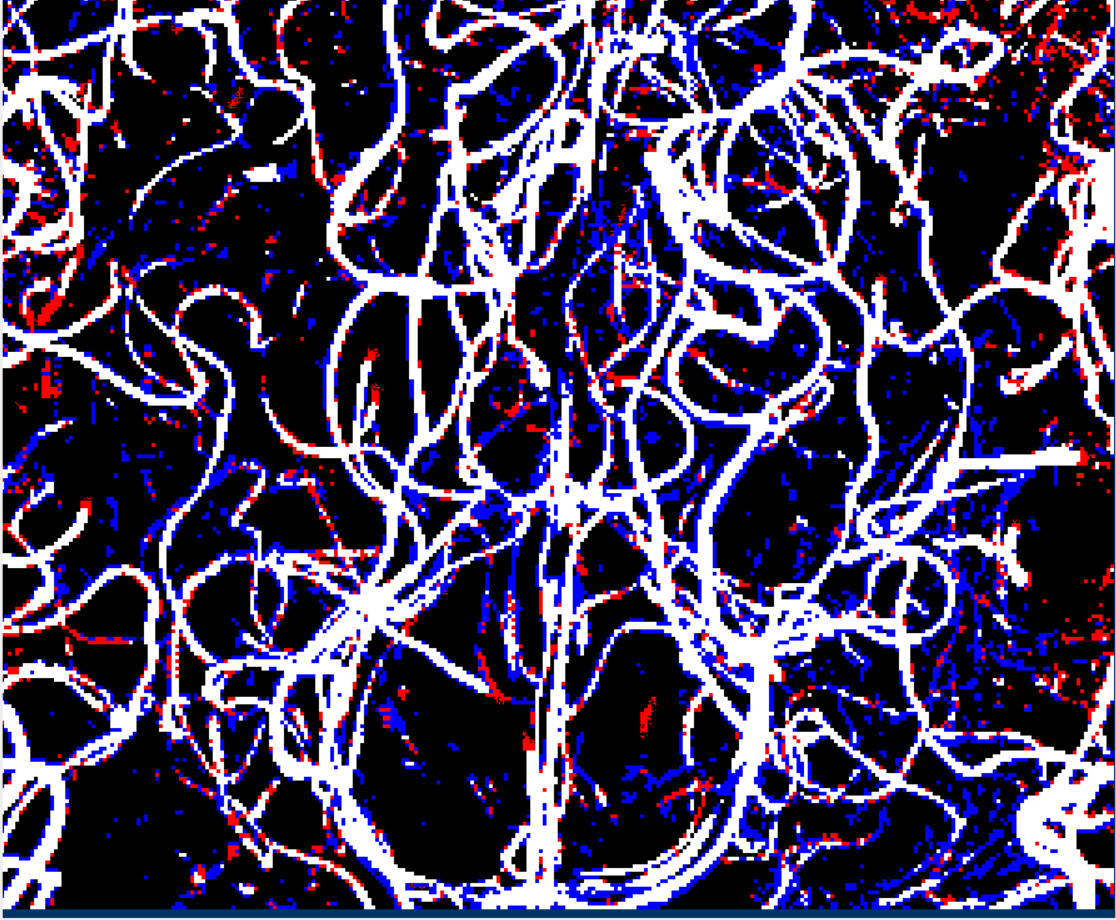} \\

    \includegraphics[angle=180,origin=c,width=.23\linewidth]{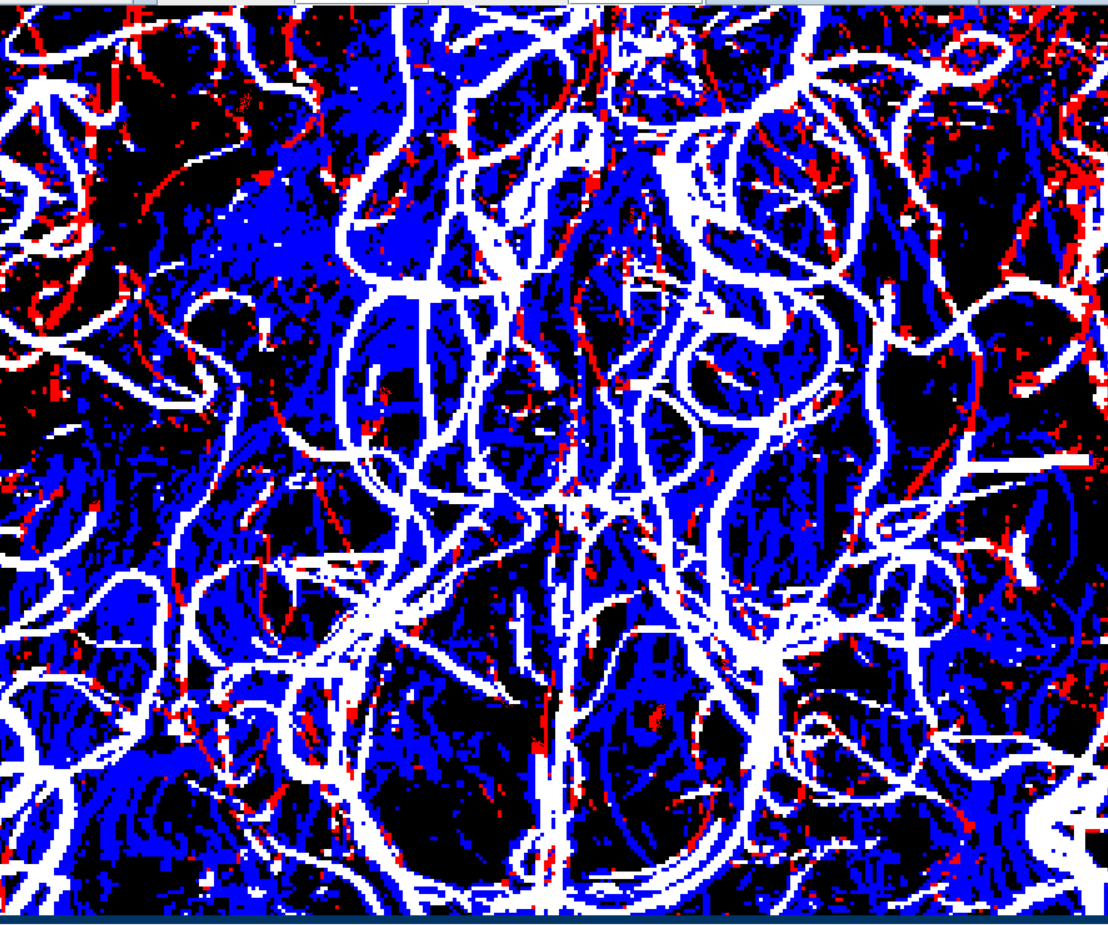} &
    \includegraphics[angle=180,origin=c,width=.23\linewidth]{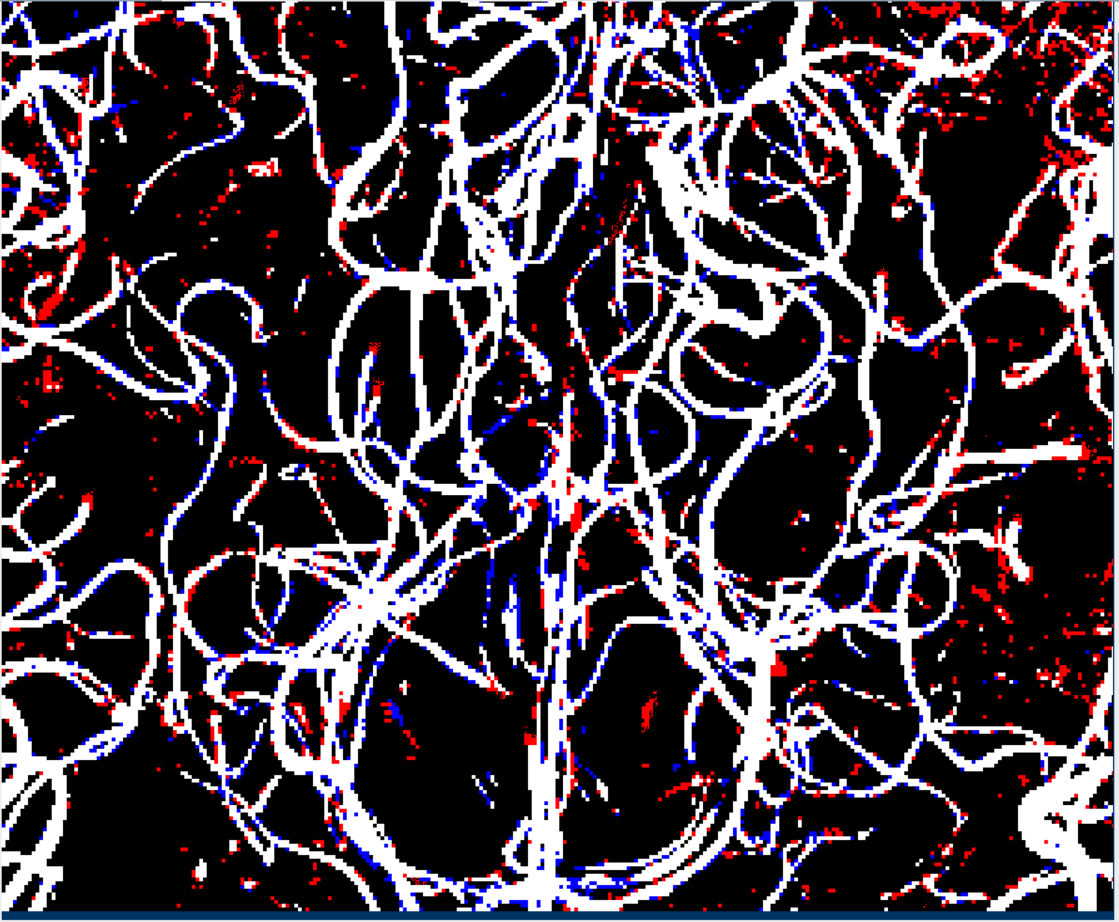} &
    \includegraphics[angle=180,origin=c,width=.23\linewidth]{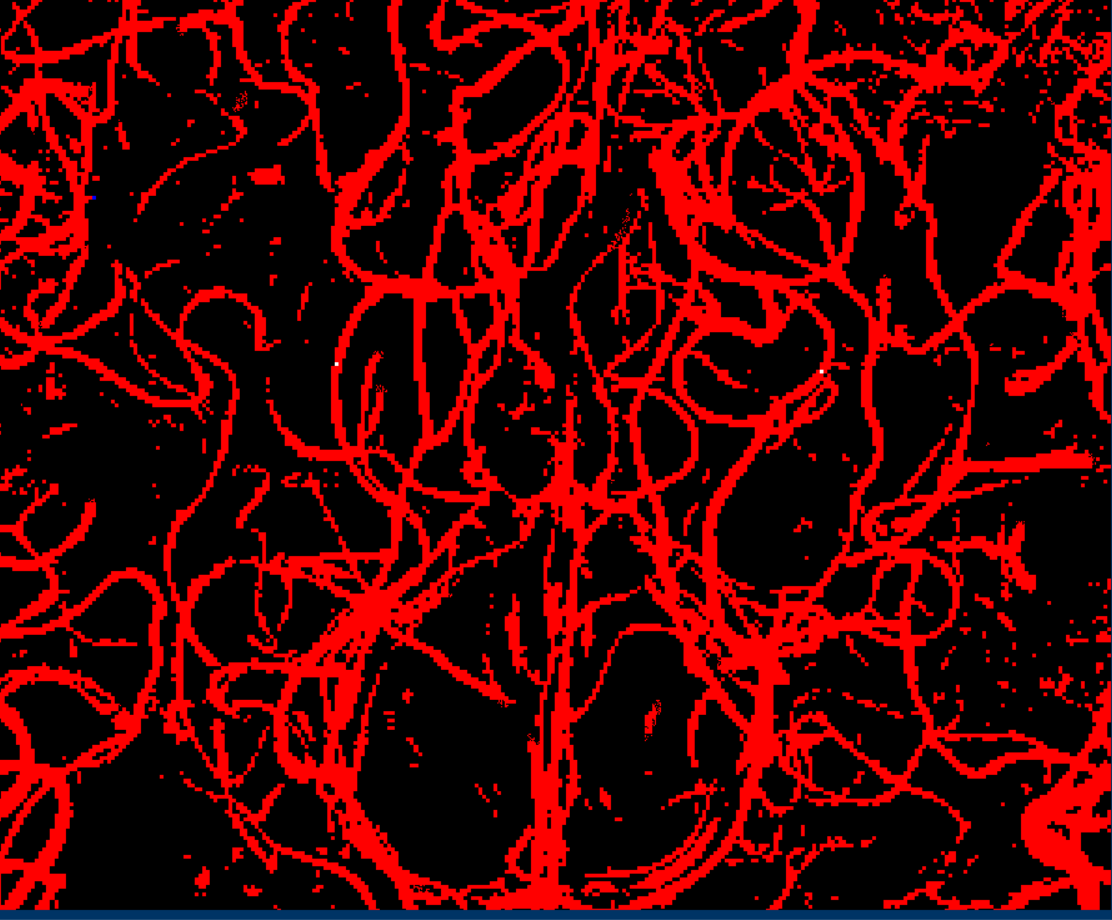} \\
   \multicolumn{3}{c}{({\bf a})}\\

    \includegraphics[angle=180,origin=c,width=.23\linewidth]{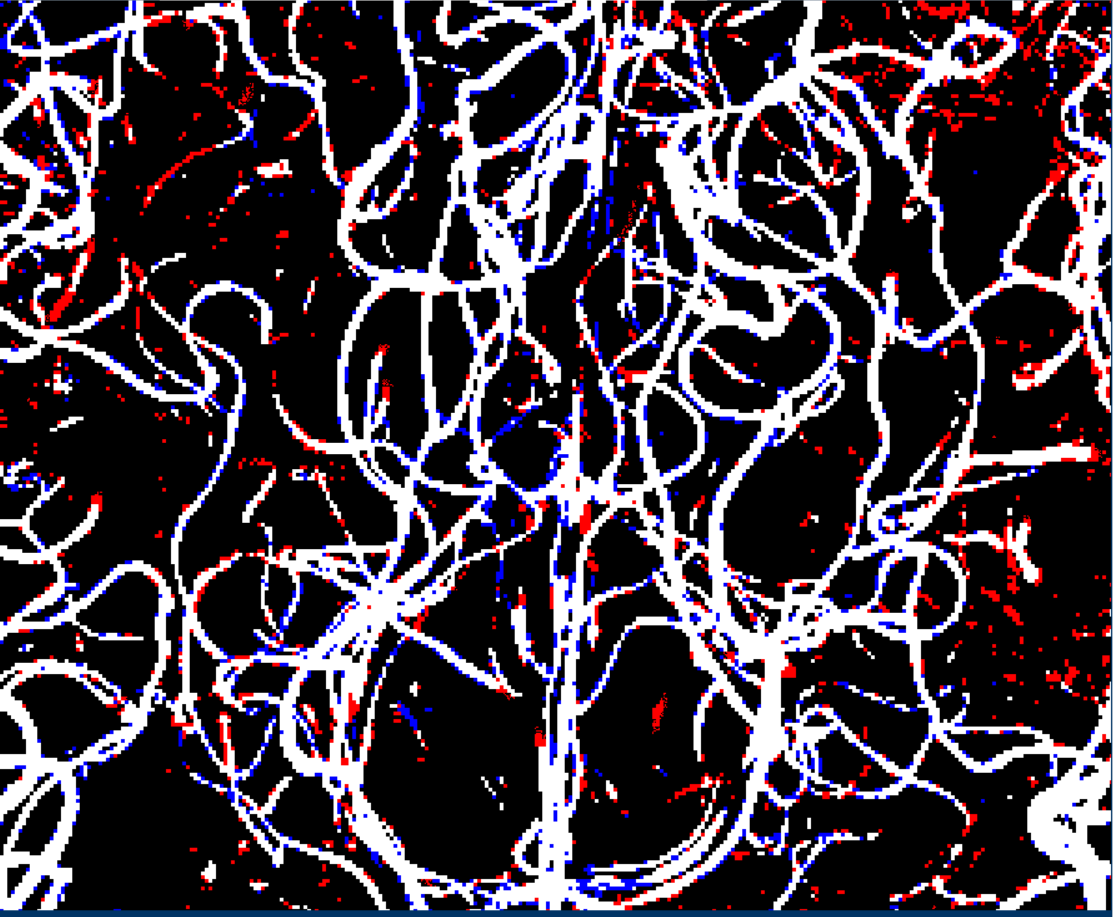} &
    \includegraphics[angle=180,origin=c,width=.23\linewidth]{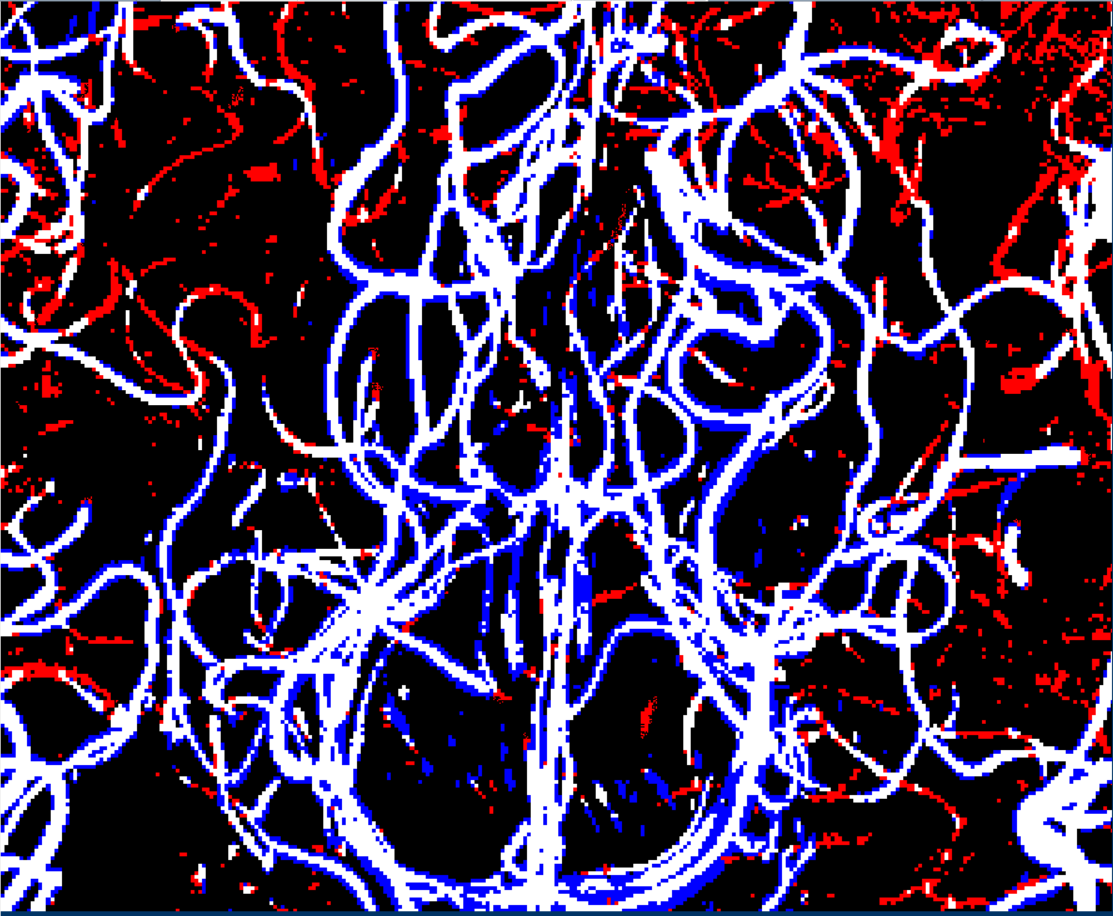} &
    \includegraphics[angle=180,origin=c,width=.23\linewidth]{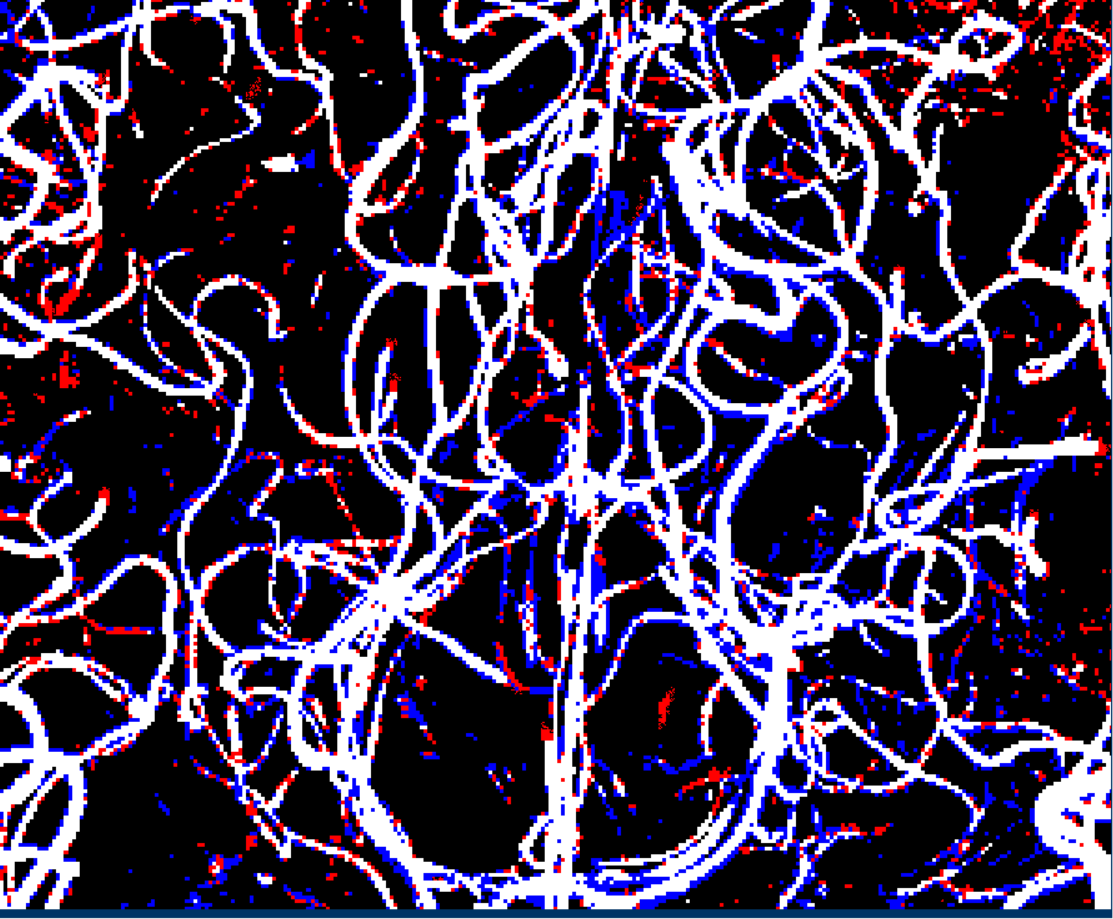} \\
  
    \includegraphics[angle=180,origin=c,width=.23\linewidth]{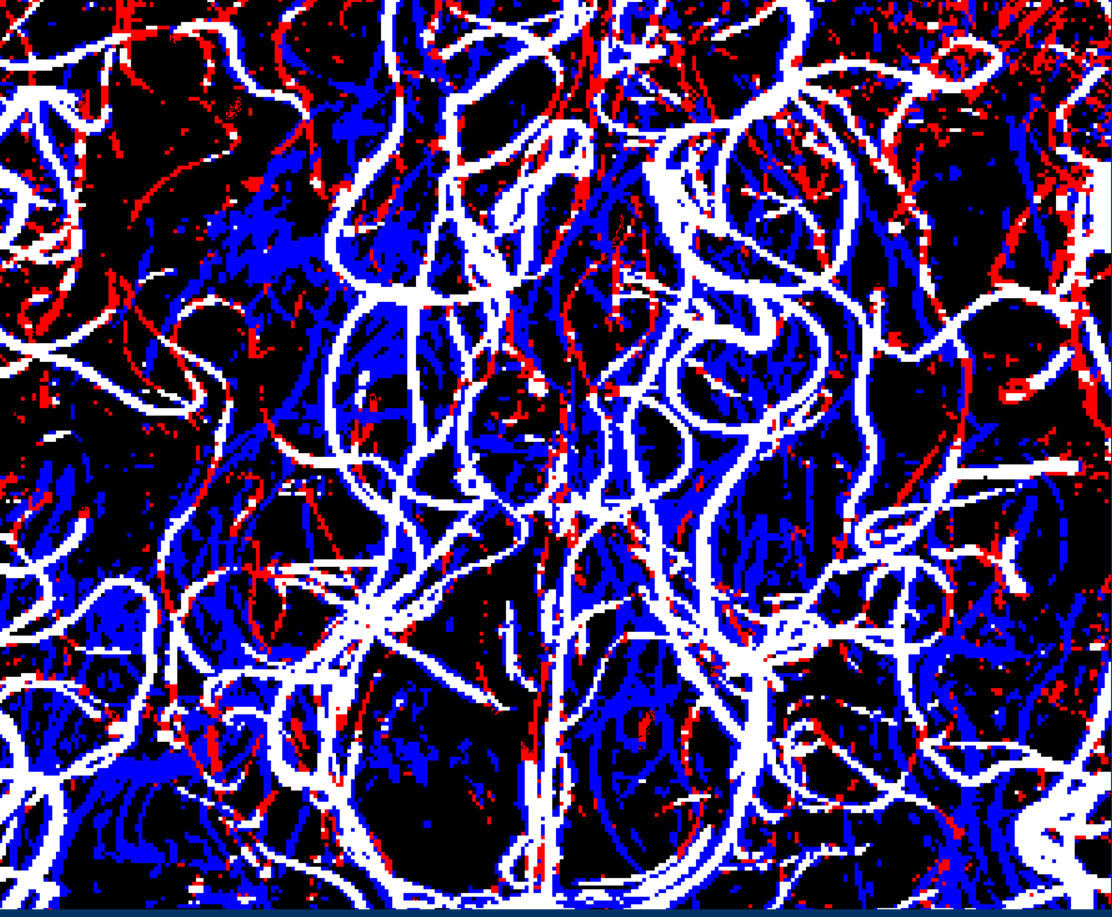} &
    \includegraphics[angle=180,origin=c,width=.23\linewidth]{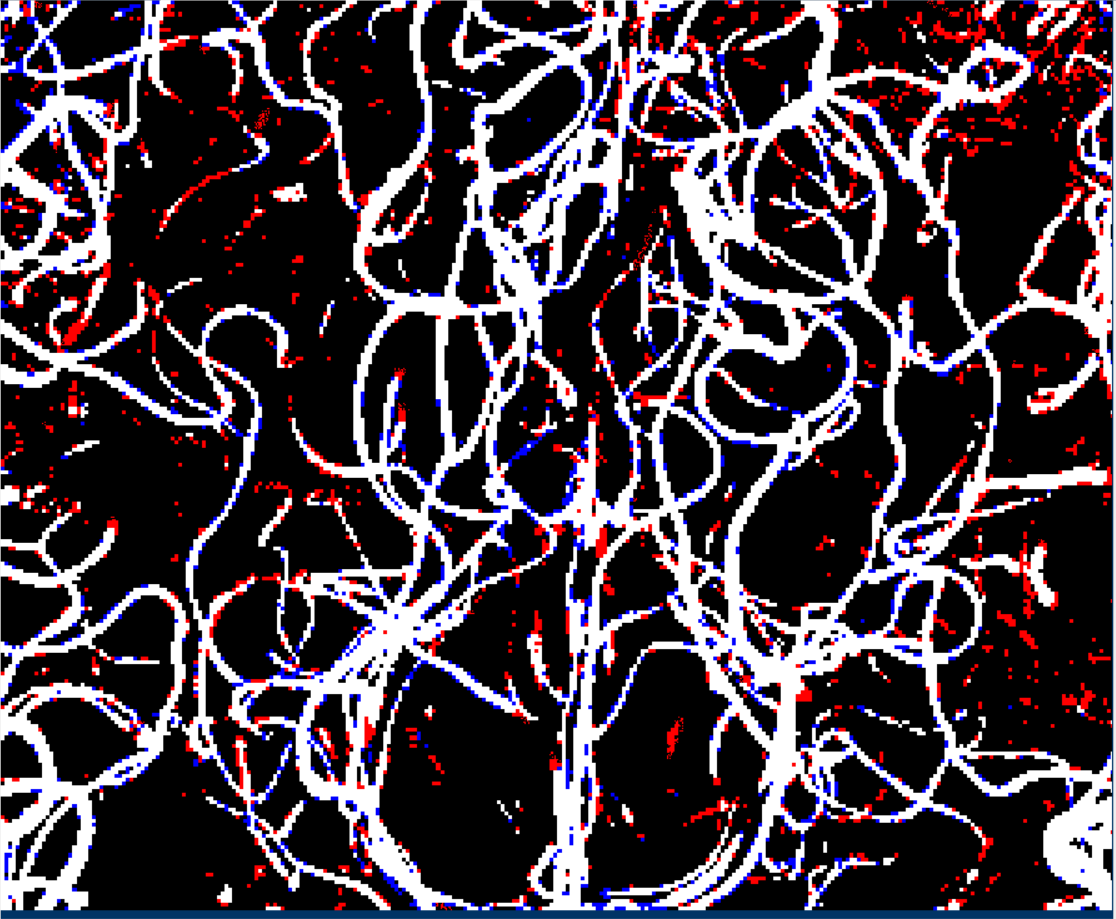} &
    \includegraphics[angle=180,origin=c,width=.23\linewidth]{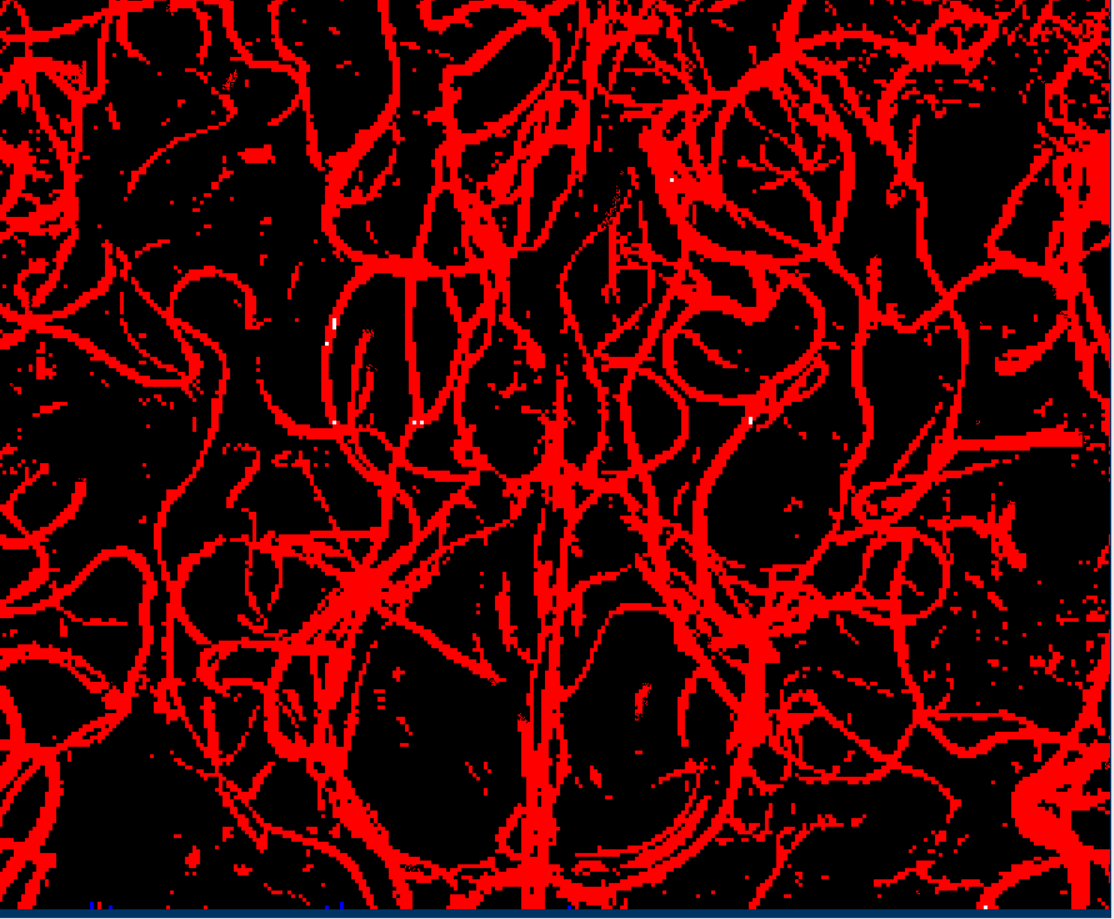} \\
     \multicolumn{3}{c}{({\bf b})}\\
  \end{tabular}

\caption{Performance 
 of the models \textit{with 
} deformation-aware learning, while dealing with various MR artefacts: (First row left to right) Random Bias Field, Random Blur, Elastic deformation (Second row left to right) Random Motion, Random Noise, and Random Spike. Red indicates false negative and blue indicates false positive while comparing against the dataset labels. (\textbf{a}) U-Net + deform. (\textbf{b}) U-Net MSS + deform.}        
\label{figA3}
\end{figure}

To further analyse the behaviour of the models, a qualitative analysis was performed on models with and without deformation, the results are reported in the Figures~\ref{figA2} and~\ref{figA3},
 respectively.
  From these figures, it can be inferred that the non-deformed models 
   perform relatively better qualitatively in terms of random spikes, but with the random motion transform, the segmentation is seen to be better in models using deformation.
\vspace{-6pt}

\begin{table}[H]
\caption{Different types of MR artefacts, simulated using TorchIO Random Transformations.}
\label{tabA1}
\begin{adjustwidth}{-\extralength}{0cm}
\newcolumntype{C}{>{\centering\arraybackslash}X}
\begin{tabularx}{\fulllength}{m{5cm}<{\raggedright}l}
\toprule
\textbf{Artefact Type}  & \textbf{Description}                             \\ 
\midrule
Random Spike 
   & Adds random stripes in different directions      \\ 
Elastic Deform & Applies dense random elastic deformation         \\
Random Noise   & Adds random Gaussian noise                      \\
Random Blur    & Blurs the image with Gaussian filter            \\
Random Motion  & Simulates motion artefacts  \\
Random Bias Field & The bias field is modelled as a linear combination of polynomial basis functions.\\
\bottomrule
\end{tabularx}
\end{adjustwidth}
\end{table}

\vspace{-10pt}

\section[\appendixname~\thesection. Mixed Precision Training]{Mixed Precision Training}
\label{app:mixedprecis}
\begin{figure}[H]
\begin{tikzpicture}
  \begin{axis}[
      ylabel      = {$Dice Coefficient$},
      xmin        = 0,
      xmax        = 12,
      domain      = -1:5
      axis lines  = center,
      xtick       = { 1,3,5,7,9,11},
      xticklabels = { U-Net(FP), U-Net(MP), AttU-Net(FP),  AttU-Net(MP),  U-Net MSS(FP),  U-Net MSS(MP), },
    x tick label style={rotate=45,anchor=east},
    ]
    
    \addplot table[x index=0, y index=1] {data/apex.dat}; 
    \addplot table[x index=2, y index=3] {data/apex.dat}; 
    \addplot table[x index=4, y index=5] {data/apex.dat}; 
    \addplot table[x index=6, y index=7] {data/apex.dat}; 
    \addplot table[x index=8, y index=9] {data/apex.dat}; 
    \addplot table[x index=10, y index=11] {data/apex.dat}; 

  \end{axis}
\end{tikzpicture}
\caption{Dot plot 
 for Models trained with full precision (FP) and mixed precision (MP) using Apex, tested with three 7T MRAs.}
\label{figA4}
\end{figure}
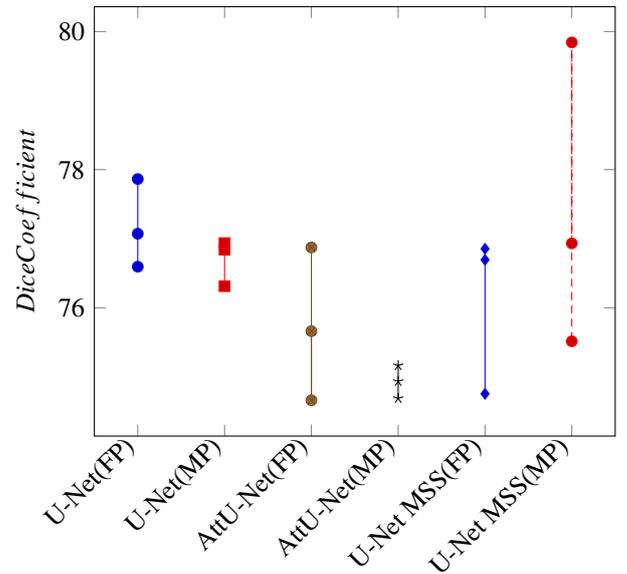

The network was trained on Nvidia Tesla V100-SXM2-32GB for 50 epochs which took 40 h. A significant decrease in memory utilisation can be observed with the mixed-precision training~\citep{micikevicius2017mixed}. Mixed precision decreases the memory consumption for the model by using half-precision values (16-bit floating-point or FP16) instead of regular full precision values (32-bit floating-point or FP32). The training setup could hold only eight 3D patches (64$^3$) with FP32, while the mixed-precision allowed the model to train a batch of 20 patches (64$^3$), which effectively reduced the time needed for each epoch to finish. Table~\ref{tabA2} shows that metrics are comparable in both cases. Figure
~\ref{figA4} is a dot plot for the same table, where the Dice coefficients of the test images compared against the respective dataset labels can be seen in each of the cases. 

\begin{table}[H]
\caption{Quantitative evaluation of full precision and mixed precision trainings of the \mbox{baseline models}.}
\label{tabA2}
\newcolumntype{C}{>{\centering\arraybackslash}X}
\begin{tabularx}{\textwidth}{CCCCC}
\toprule
{} &  \multicolumn{2}{c}{\textbf{Full Precision}} & \multicolumn{2}{c}{\textbf{Mixed Precision}}\\

\midrule
\textbf{Model} & \textbf{Dice Coefficient} & \textbf{IoU} &  \textbf{Dice Coefficient} & \textbf{IoU}\\
\midrule
U-Net  &  {77.22} & {62.89} &  {76.93} & {62.51}\\
\midrule
Attention U-Net & {75.84} & {61.08} &  {76.53} & {61.98}\\
\midrule
U-Net MSS & {76.08} & {61.40} &  {77.42} & {63.00} \\
\bottomrule
\end{tabularx}
\end{table}

\begin{adjustwidth}{-\extralength}{0cm}

\reftitle{References}

\end{adjustwidth}
\end{document}